\documentclass[a4paper]{spie}  
\usepackage{txfonts}
\usepackage{bm}
\usepackage[latin1]{inputenc}
\usepackage[francais]{babel}
\usepackage{rotating} 
\usepackage{txfonts}
\usepackage{color}

\title{Direct imaging with a hypertelescope of red supergiant stellar surfaces} 

\author{Patru F.\supit{a}, Chiavassa A.\supit{b}, Mourard D.\supit{c} and Tarmoul N\supit{c}
\skiplinehalf
\supit{a}European Southern Observatory, Alonso de Cordova 3107, Vitacura, Santiago, Chile; \\
\supit{b}Max Planck Institute for Astrophysics Germany, Karl-Schwarzschild-Str. 1, 85741 Garching, Germany; \\
\supit{c}Laboratoire Fizeau UMR6525, Avenue Nicolas Copernic, 06130 Grasse, France;
}

\authorinfo{Further author information: Send correspondence to Fabien
  Patru (E-mail: fpatru@eso.org) or Andrea Chiavassa (E-mail: chiavass@mpa-garching.mpg.de)\\}

\begin{document} 
  \maketitle 

\begin{abstract}
High angular resolution images obtained with a hypertelescope can strongly constrain the radiative-hydrodynamics simulations of red supergiant (RSG) stars, in terms of intensity contrast, granulation size and temporal variations of the convective motions that are visible on their surface. The characterization of the convective pattern in RSGs is crucial to solve the mass-loss mechanism which contributes heavily to the chemical enrichment of the Galaxy.
We show here how the astrophysical objectives and the array configuration are highly dependent to design a hypertelescope.
For a given field of view and a given resolution, there is a trade-off between the array geometry and the number of required telescopes to optimize either the (u,v) coverage (to recover the intensity distribution) or the dynamic range (to recover the intensity contrast).
To obtain direct snapshot images of Betelgeuse with a hypertelescope, a regular and uniform layout of telescopes is the best array configuration to recover the intensity contrast and the distribution of both large and small granulation cells, but it requires a huge number of telescopes (several hundreds or thousands). An annular configuration allows a reasonable number of telescopes (lower than one hundred) to recover the spatial structures but it provides a low-contrast image.
Concerning the design of a pupil densifier to combine all the beams, the photometric fluctuations are not critical ($\Delta$ photometry $<50\%$) contrary to the residual piston requirements (OPD $<\lambda/8$) which requires the development of an efficient cophasing system to fully exploit the imaging capability of a hypertelecope.
\end{abstract}



\section{INTRODUCTION}

Current ground-based optical interferometers are already able to provide images, using the model fitting and the aperture synthesis techniques \cite{Perrin2007,Kraus2007,LeBouquin2009,Lacour2009,Chiavassa2010}. However, these images are limited both in spatial resolution and in spatial frequency coverage, so that the image reconstruction is always an undetermined problem and requires a pragmatic method to be achieved. Taking into account the astrophysical objectives in high angular resolution \cite{Lena2004}, the future interferometric arrays should have many telescopes with kilometric baselines to enhance their imaging capabilities. They also require a suitable beam-combiner to manage the high number of entrance sub-apertures. Current pair-wise beam-combiner are not suitable in that case, and an all-in-one combination scheme should be used. 

The hypertelescope concept appears as a convenient and efficient solution \cite{Labeyrie1996,Labeyrie2007,Labeyrie2008}. Using many telescopes, an efficient cophasing system and a densified pupil combiner, they can provide direct snaphot images of complex targets. Due to the highly diluted entrance pupil of the telescopes spread on several hundred meters or kilometers, a multi-axial combination leads to a faint and spread Fizeau image formed in the common focal plane. A pupil densifier rearrange the beams so as to preserve the high resolution information and to highly intensify the signal regarding to the Fizeau mode. By increasing the relative size of the beams and keeping the general layout of the position of the beams, most of the light is concentrated in the useful field of view (FOV) \cite{Lardiere2007}, reducing the image width to a convenient size for the detectors whatever the lengths of the baselines. The characteristics of the point spread function (PSF) have already been described in a previous paper \cite{Patru2009}, showing that there is a trade-off between the angular resolution, the useful field of view and the halo level of the PSF to choose an optimal array configuration. This paper deals with the design of an array and of a combiner in the case of stellar surface imaging applications.

We concentrate here on the case of red supergiant (RSG) stars. Massive stars with masses between roughly 10 and $25~M_{\odot}$ spend some time as RSGs being the largest stars in the universe. They have effective temperature ($T_{\rm eff}$) ranging from 3450 to $4100~K$), luminosities of 20\,000 to 300\,000 $L_{\odot}$, and radii up to 1500 $R_{\odot}$ \cite{2005ApJ...628..973L}, which makes them some of the brightest stars known. Such extreme properties foretell the demise of a short-lived stellar king because they are nearing the end of their life and they are doomed to explode as a supernova. RSGs still hold several unsolved mysteries: (i) the mass-loss mechanism, shedding tremendous quantities of gas, is unidentified; (ii) their chemical composition is largely unknown because of difficulties in analyzing their complex spectra due to the low surface temperatures and vigorous convection. RSGs contribute extensively to the chemical enrichment of our Galaxy loosing enormous quantities of their mass due to an unknown process. The vigorous convection that they experience could be at the base of the mass-loss mechanism.

Recently, 3D radiative-hydrodynamics (RHD) simulations done with the numerical code CO$^5$BOLD \cite{2002AN....323..213F, Freytag2003SPIE.4838..348F,Freytag2008A&A...483..571F} coupled with the calculations done with radiative transfer code OPTIM3D \cite{2009A&A...506.1351C,2010A&A...515A..12C} have reported robust interferometric comparisons of hydrodynamical simulations with existing observations \cite{2000MNRAS.315..635Y,2004young,2004A&A...418..675P,2009A&A...508..923H} in the optical, $H$ and $K$ band regions, arguing for the presence of convective cells of various sizes on the red supergiant Betelgeuse. The characterization of the convective pattern in RSGs is crucial to solve the mass-loss problem and high angular resolution imaging with hypertelescopes will constrain the RHD simulations in terms of surface intensity contrast, granulation size and temporal variations. \\

\section{Astrophysical requirements in the case of Betelgeuse} \label{}

\subsection{Astrophysical parameters used for radiative-hydrodynamics (RHD) simulations} \label{}

\begin{figure}[!h] 
\begin{center}
\begin{tabular}{cccc}
\includegraphics[width=50mm]{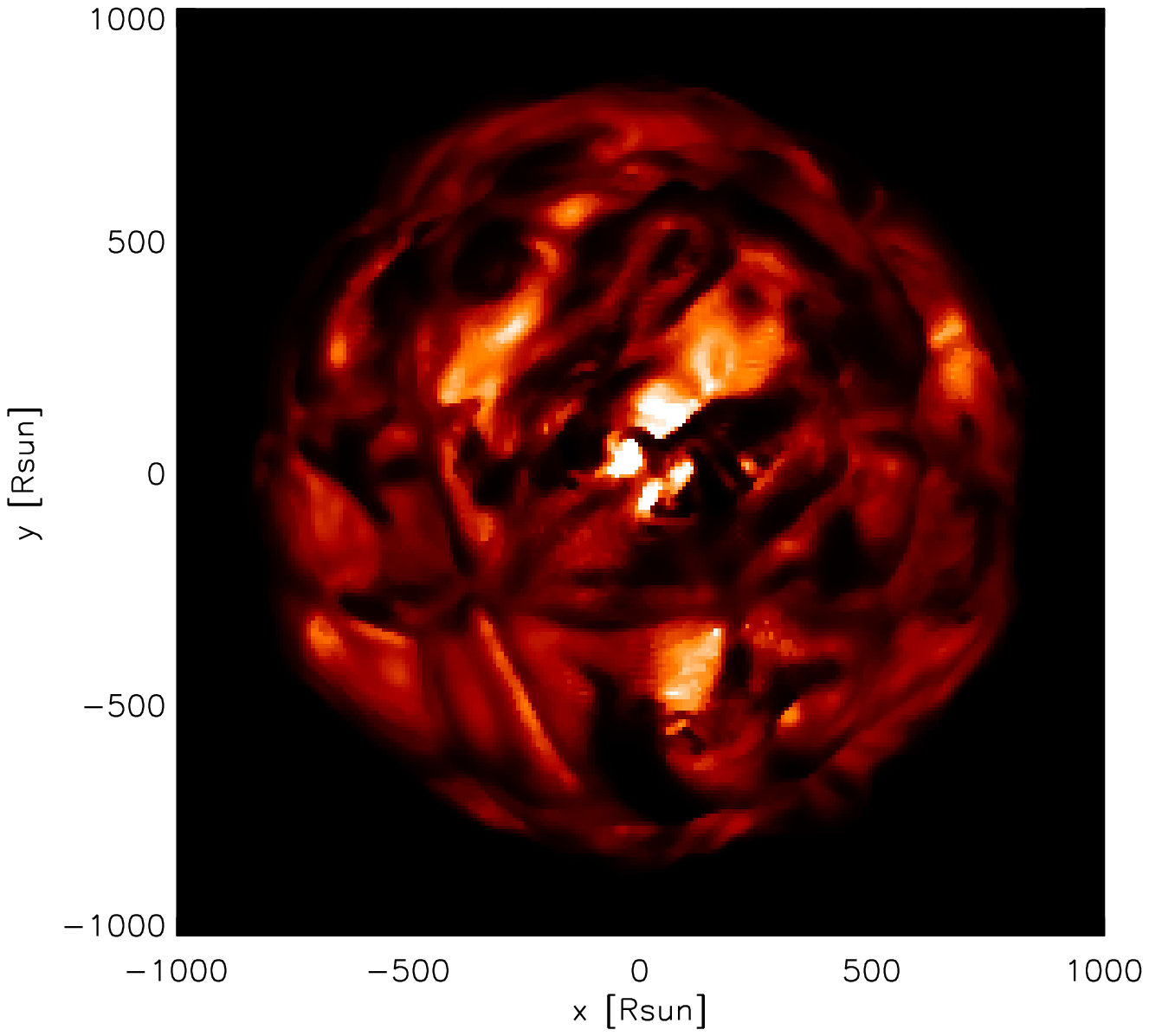} &
\includegraphics[width=50mm]{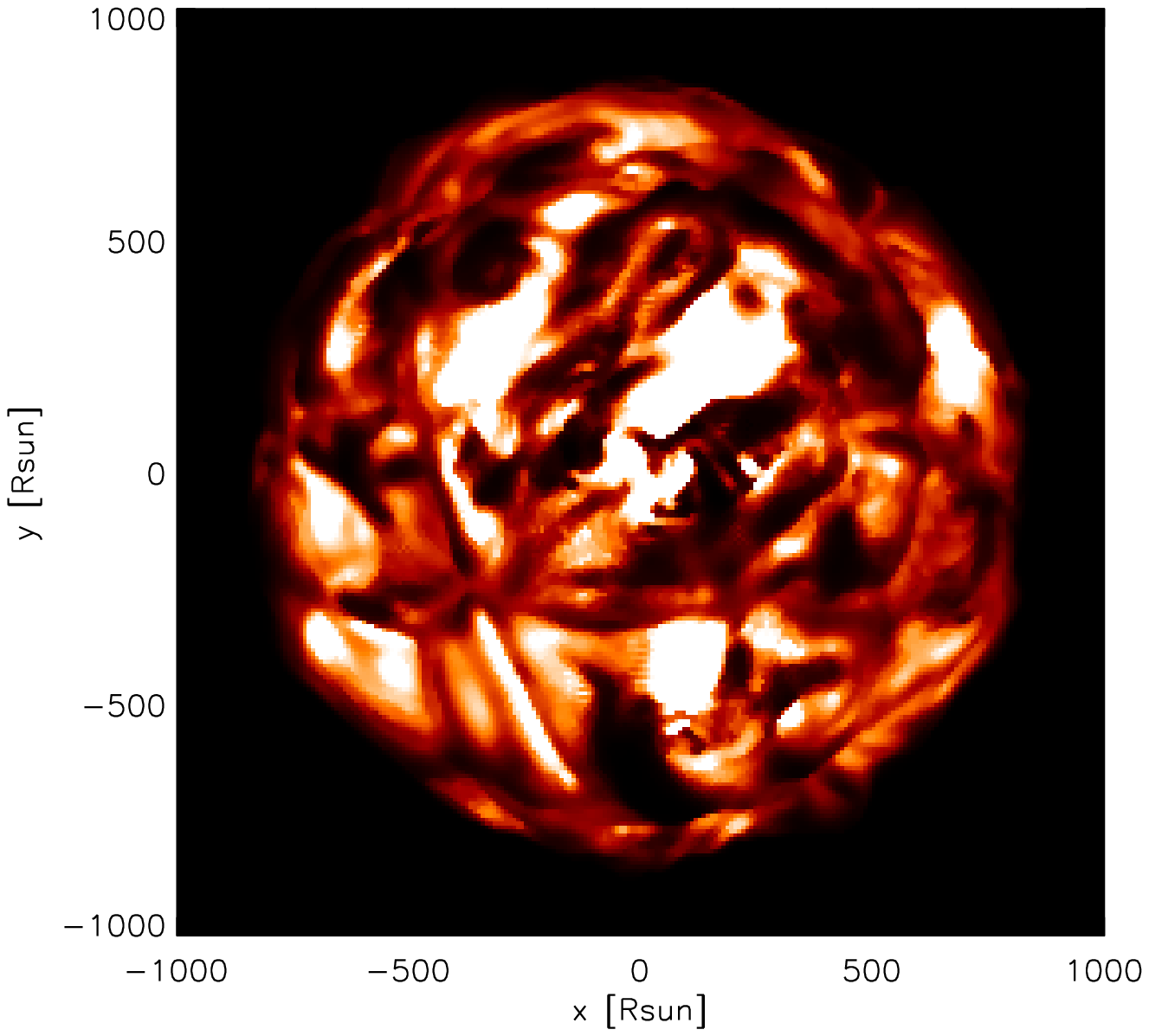} &
\includegraphics[height=40mm]{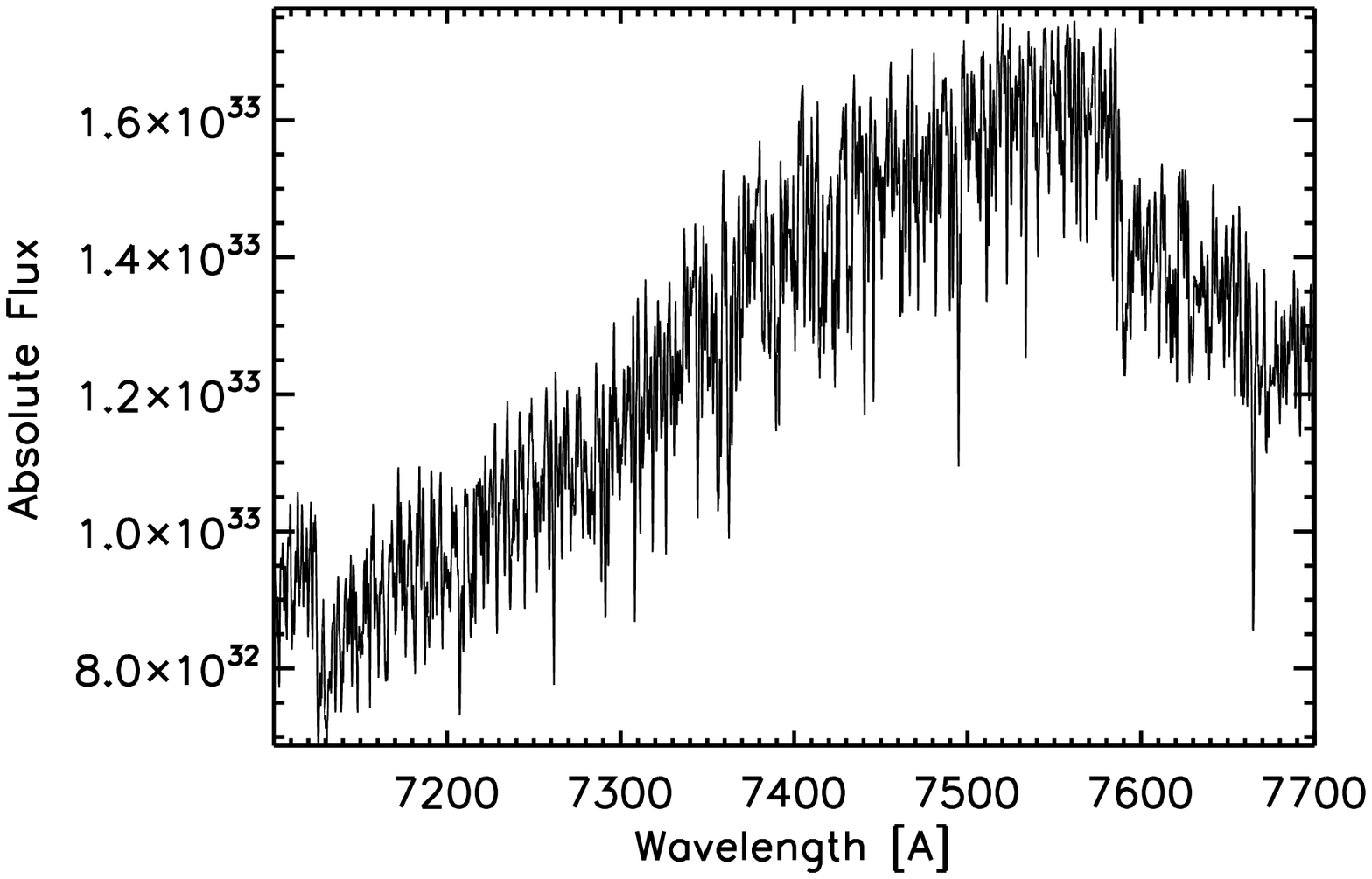} \\
\end{tabular}
\end{center}
\caption[]
{Original map of the linear intensity of Betelgeuse at 7450 \AA, with 2 different scale to underline the structures, as the intensity covers a large dynamic range. The range is [0,600 000] erg/s/cm$^2$/\AA (left) and [0,300 000] erg/s/cm$^2$/\AA (middle), whereas the maximum intensity of the central bright cell reaches 1200 000 erg/s/cm$^2$/\AA. Synthetic spectrum of Betelgeuse on the range [7100\AA-7800\AA] (right).
\label{originalBetelgeuse} }
\end{figure} 

Actual RHD simulations aim to reproduce the stellar parameters of the prototypical RSG star Betelgeuse (HD~39801,
M1--2Ia--Ibe), which is one of the best studied RSGs in term of multi-wavelength imaging because of its large luminosity and angular diameter. The model used in this work has the closest stellar parameters corresponding to this star : a 12 $M_{\odot}$ stellar mass, a numerical resolution of $235^3$ grid points with a step of 8.6~$R_{\odot}$, a luminosity of $L=93\,000\pm1300~L_{\odot}$ averaged over the spherical shells and over time, an effective temperature of $T_{\rm{eff}}=3490\pm13~K$ (compared to $T_{\rm{eff}}=3640~K$ \,\cite{2005ApJ...628..973L}), a radius of $R=832\pm0.7~R_{\odot}$, and a surface gravity $log(g)=-0.337\pm0.001$ (compared to $log(g)=-0.3$ \,\cite{2008AJ....135.1430H}).

We have computed intensity maps using OPTIM3D for the RHD simulation at different wavelengths in the region between 710 and $780~nm$ (in total 700 images, $0.1~nm$ apart). This spectral region is interested by strong molecular absorption (namely TiO; Fig. \ref{originalBetelgeuse}) that cause high-contrast patterns characterized by dark spots and bright areas. The brightest areas exhibit an intensity 50 times brighter than the dark ones with strong changes over some weeks \cite{2010A&A...515A..12C}, as shown in Fig.~\ref{originalBetelgeuse}. 
\\

\subsection{Astrophysical parameters for spectral analysis} \label{wave}

Three different spectral resolutions would be interesting to explore:

\begin{itemize}
\item Low resolution ($R\approx 5$) to study the large scale granule. This is more interesting in the $H$ and $K$ bands where only few convective cells of the size of $\approx 60\%$ of the stellar radius cover the whole surface of the simulated RSG \cite{2009A&A...506.1351C}.
\item Medium resolution ($R\approx 20-30$) to study the temperature stratification in the atmosphere. Observations at wavelengths in a molecular band and in the close pseudo-continuum probe different atmospheric depths, and thus layers at different temperatures that highly constraint the limb-darkening effect of RHD simulations \cite{2010A&A...515A..12C}.
\item High resolution ($R\approx 1000$ or more) to study the cinematic in the spectral lines. In fact, these stars have velocities up to $20-30~km/s$ in the optical thin region that strongly affect the line broadening. 
\end{itemize}

In the following analysis of this paper, we consider only the intensity map at $745~nm$ (Fig. \ref{originalBetelgeuse}).
\\
\\

\subsection{Objectives to image Betelgeuse} \label{}

The main objectives that we need to recover from future interferometric images in order to better constrain the simulations are:
\begin{itemize}
\item[1.] the size of the granulation cells from the large scale granule to the small-medium scale convective-related structures;
\item[2.] the surface intensity contrast;
\item[3.] the timescale of the convective motions. 
\\
\end{itemize}

These objectives are crucial :
\begin{itemize}
\item to understand the photosphere and dynamics of RSGs where about half of the atomic elements are produced;
\item to explain the (unknown) mass-loss process which releases material to the surrounding medium and contributes
heavily to the chemical enrichment of the Galaxy;
\item to determine accurate abundances in distant galaxies from complex spectra by using proper model atmospheres.
\\
\end{itemize}

A previous attempt has been done \cite{Berger2010} to review the status of current or soon-to-be interferometers in terms of image synthesis. Interferometric data have been simulated from synthetic images in the $K$ band (computed with the same image as in Fig. \ref{originalBetelgeuse} and the same numerical codes used for this work). Reconstructed images of RSGs have been provided with cutting-edge code using the simulated data obtained with the VLTI facilities in a 6-8 telescopes configuration and with the earth rotation to provide a large (u,v) coverage.

The reconstructed image of Betelgeuse shows in that case medium sized granules, similar to the original image smoothed to the interferometer resolution. There is, however, much more information in the original image (Fig.  \ref{originalBetelgeuse}). Key features like the large central granule are very hard to see and the narrow inter-granular lanes are not seen at this angular resolution. 

Moreover, it appears that there is not apparent correlation of the intensity contrast between the smoothed and the reconstructed images. Then, the limitations of current image-reconstruction techniques lie in their ability to match exact intensities and to detect highly contrasted features. Indeed, the dynamic range is a function of the number of (u, v) points and of the accuracy of the visibility measurements \cite{Berger2010}. 

The timescale of the convective pattern is also crucial to constrain the simulations of red supergiant stars. A follow-up study at different epochs is probably needed at maximum 1 month apart in the optical wavelength where things change faster on timescales of days or even hours. This requires snapshot images and/or as many simultaneous measurements as possible in a short period \cite{Berger2010}.

It has been concluded \cite{Berger2010} that the complex structure of Betelgeuse requires a rich (u,v) plane sampling with a higher resolution and therefore the combination of the highest number of telescopes and of larger baselines. A good accuracy of the interferometric measurements is also mandatory by using for instance spatial filtering features and efficient fringe tracking systems.\\
\\
\\
\\

\subsection{Correlation criterium between images} \label{}

In this paper, we want to compare different images obtained with different array configurations and for different conditions of observations, so as to estimate the technical specifications of a hypertelescope to preserve an image of quality, regarding the astrophysical objectives of Betelgeuse.
In a first approch, we have chosen as criterium the Pearson correlation coefficient \cite{Pearson1920}, which allows to compare each image with the original one (Fig. \ref{originalBetelgeuse}). 

$$Pearson=\frac{\sum_{i=1}^{N_p} (X_i - \bar{X}) (Y_i - \bar{Y})}{\sqrt{\sum_{i=1}^{N_p} (X_i - \bar{X})^2} \sqrt{\sum_{i=1}^{N_p} (Y_i - \bar{Y})^2}}$$

where $N_p$ is the number of points in each image, $X_i$ (resp. $Y_i$) is the intensity of the $i^{th}$ point of the first (resp. second) image and $\bar{X}$ is the average of the $N_p$ intensities $X_i$.
This criterium equals to 1 in the best case (same images) and decreases down to 0 in the worst case (no correlation between the two images). In a forthcoming paper, we want to define a more relevant criterium to estimate the restitution of both spatial and amplitude intensity of the structures in the image. \\
\\
\\
\\

\section{Imaging properties in the case of Betelgeuse} \label{}

\begin{figure}[!t] 
\begin{center}
\begin{tabular}{cccc}
\includegraphics[width=39mm]{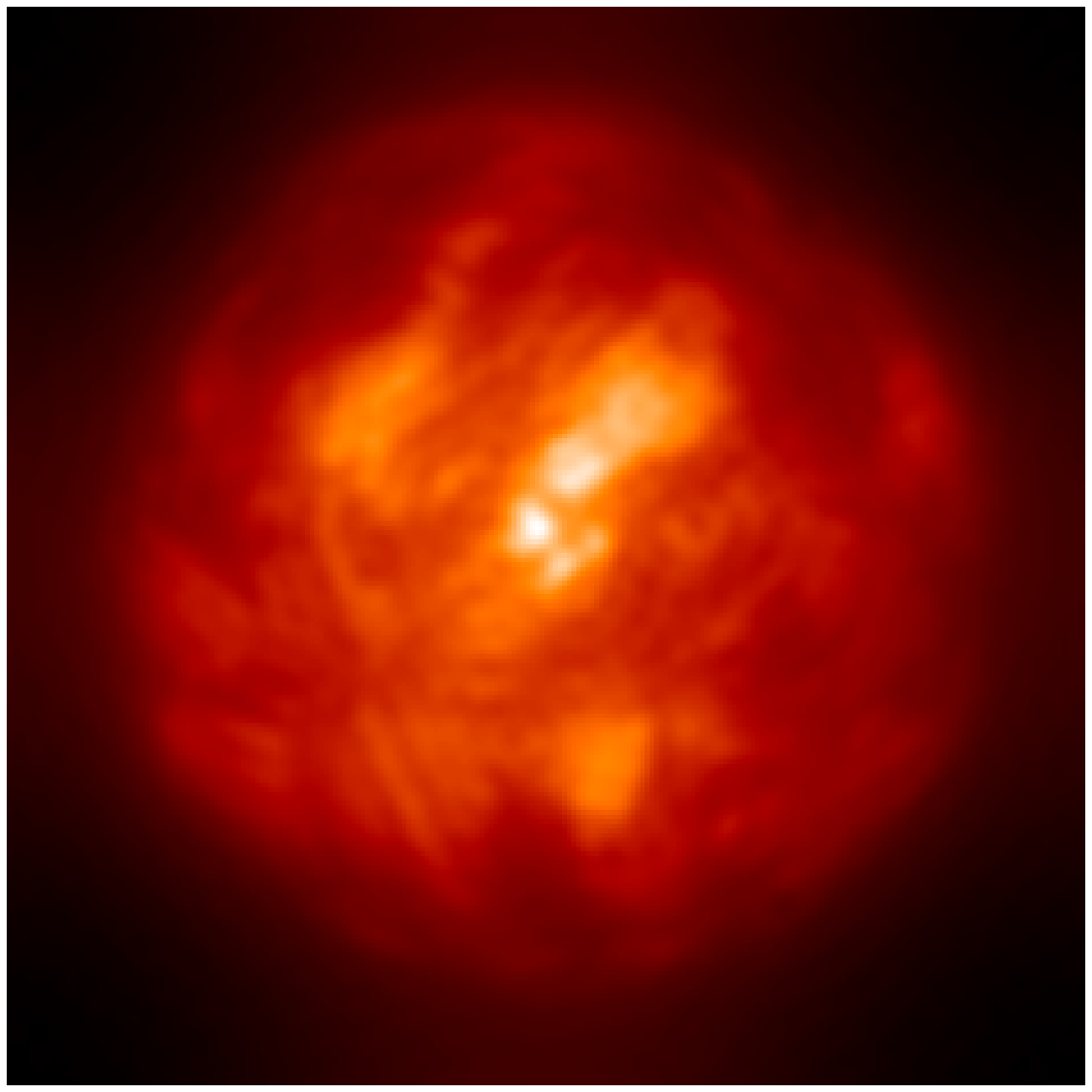} &
\includegraphics[width=39mm]{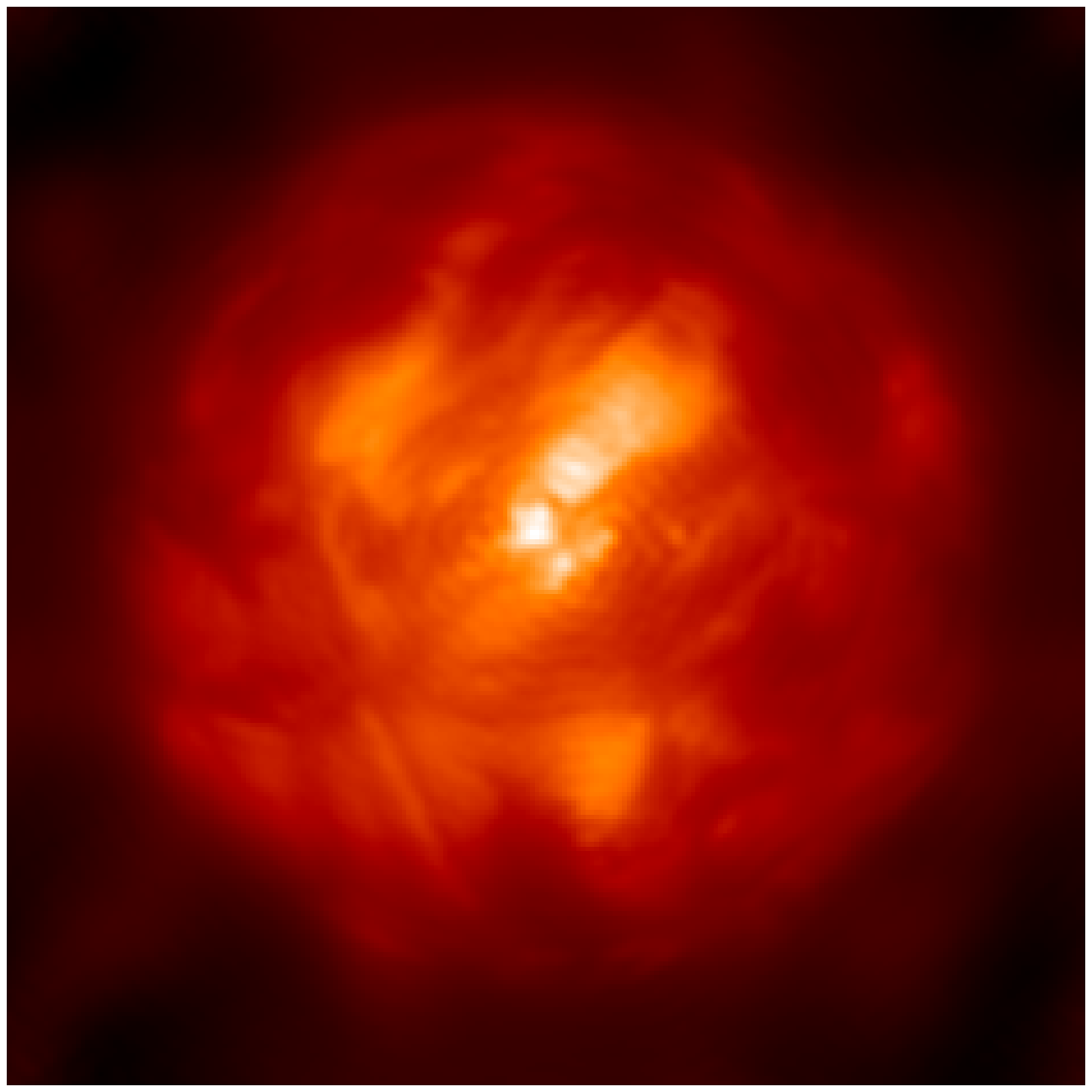} &
\includegraphics[width=39mm]{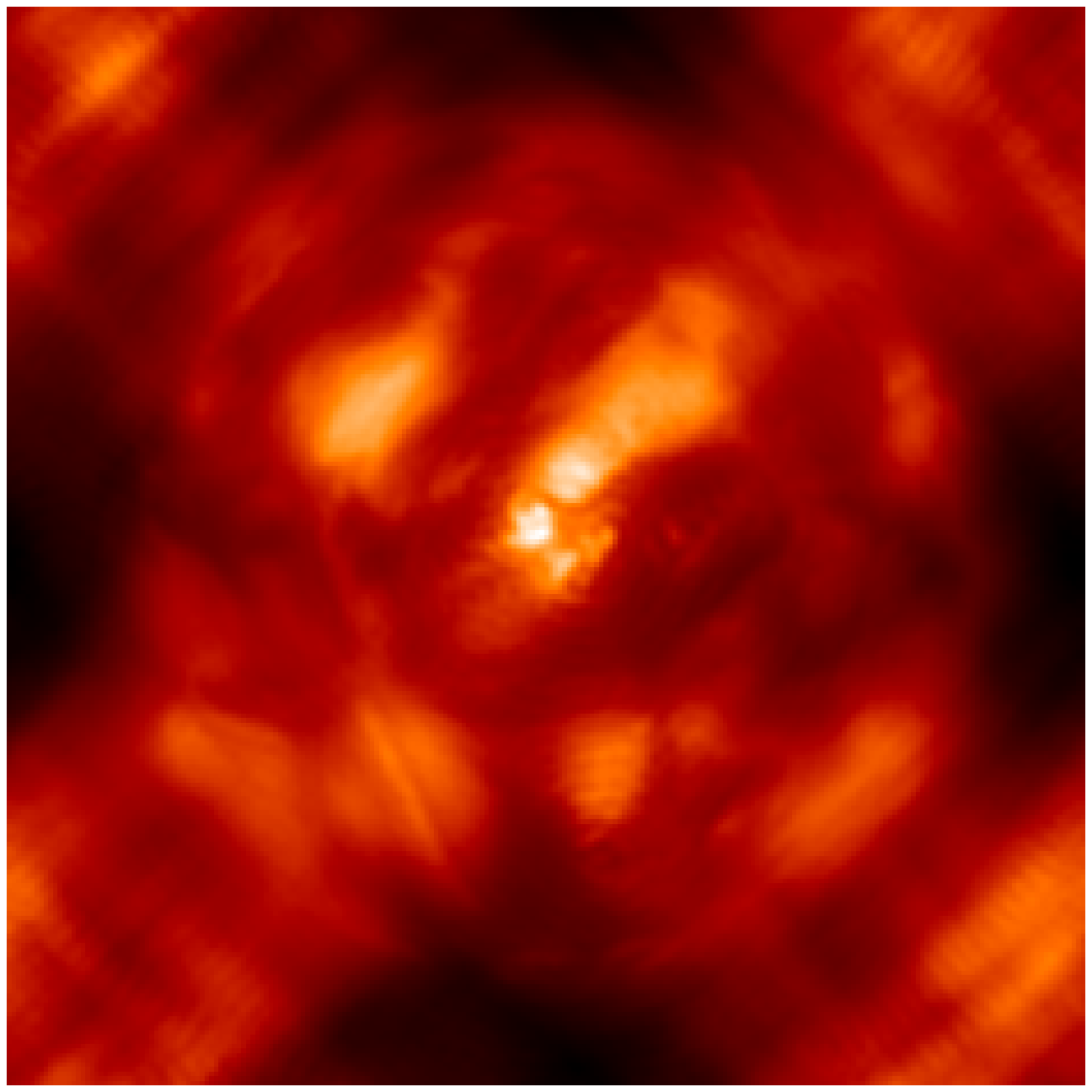} &
\includegraphics[width=39mm]{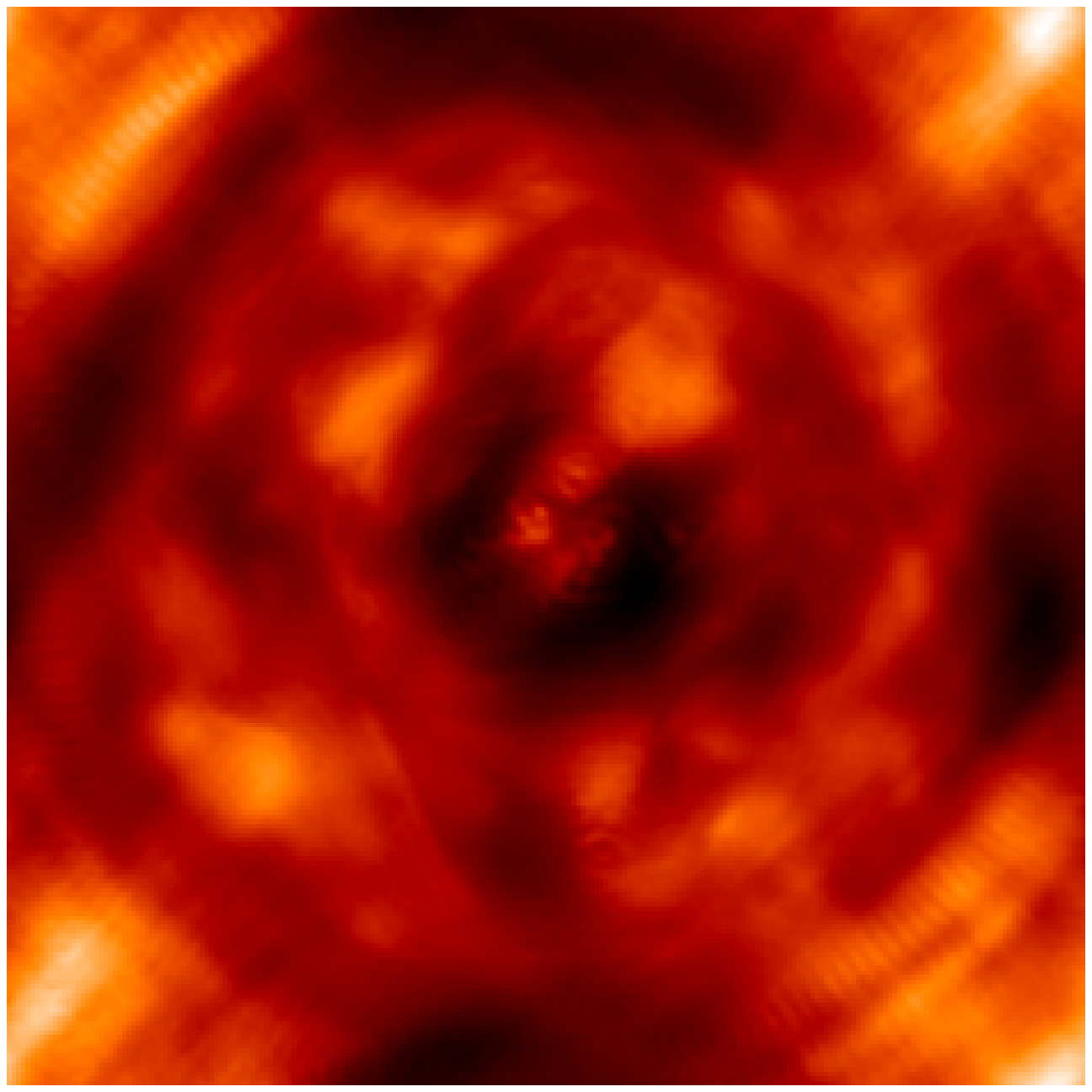} \\
\end{tabular}
\end{center}
\caption[] 
{Direct images of Betelgeuse with an annular array of 100 telescopes showing the crowding effect, which depends on the size of the clean field of view ($CLF$) compared to the object diameter ($D_*$): $D_*=1\,x\,CLF$, $D_*=1.5\,x\,CLF$, $D_*=1.8\,x\,CLF$ and $D_*=2\,x\,CLF$ (from left to right).\\
\label{OVLA100_CLFvsDstar} }
\end{figure} 

\begin{figure}[!t] 
\begin{center}
\begin{tabular}{cccc}
\includegraphics[height=60mm]{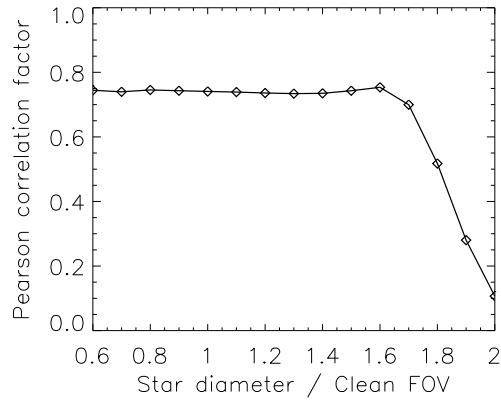}
\end{tabular}
\end{center}
\caption[] 
{Pearson correlation factor as a function of the ratio (star diameter)/(Clean FOV) to illustrate the crowding effect.\\
\\
\label{var_correl_OVLA100_CLFvsDstar} }
\end{figure} 

\subsection{The aliasing effect and the crowding effect} \label{}

The spatial frequency coverage of a given array of telescopes is directly related to its imaging properties in the direct image plan.
The highest spatial frequency or the maximum baseline $B_{max}$ corresponds to the smallest element of resolution in the image, called the $Resel$, with $Resel=\lambda/B_{max}$.
The lowest spatial frequency or the minimum baseline $B_{min}$ superimposes the lowest resolution, i.e. the largest imaging field of view, called the CLean Field of view : $CLF=\lambda/B_{min}$, where an image can be formed properly. 
We can then estimate the number of $Resels$ in the Clean FOV, which equals to $N_R=(CLF/Resel)^2=(B_{max}/B_{min})^2$.
In other words, the width of the Clean FOV equals to $CLF=B_{max}/B_{min}$ expressed in number of $Resels$.

In addition, the diameter of each telescope $D_i$ defines the portion of the sky seen by the interferometer, called the Coupled Field of view: $CF=\lambda/D_i$. A source located inside the $CF$ (but outside the $CLF$) will be "coupled" to the interferometer by forming a halo of side-lobes in the detector plane. These side-lobes induce photometric perturbations locally
distributed in the image. This phenomenon is known as the aliasing effect \cite{Aime2008,Patru2009}. 

A consequence of the aliasing effect is the crowding effect. The crowding limit is defined as the maximum number of point-like sources $N_S$ which can be contained inside the $CLF$ : $N_S<(B_{max}/B_{min})^2$ as defined above. Consequently, if the number of point-like sources inside the Coupled FOV (and inside the $CLF$) is larger than the number of $Resels$ inside the $CLF$, the image inside the $CLF$ will be saturated due to all the aliased sources.
In the case of an extended source such as Betelgeuse, the crowding appears systematically if the star diameter exceeds the $CLF$ width. Indeed, as the $CLF$ is already completely filled by the central part of the star, the edge of the star outside the $CLF$ will be aliased and will destroy the image inside the $CLF$.

Figure \ref{OVLA100_CLFvsDstar} shows the direct images obtained with an annular array of 100 telescopes, which provides the same resolution in all directions on the sky, thanks to its circular symmetry.
This figure illustrates the fact that it does not make sense to image a target that is substantially larger than the Clean FOV. This is confirmed by the figure \ref{var_correl_OVLA100_CLFvsDstar}, which shows that the Pearson correlation factor falls down above a certain value. Note that this value is not equal to 1, firstly because 1 corresponds to the width of the image whereas the diameter of Betelgeuse is quite smaller, and secondly because the edge of the star is less bright than the center.
Anyway, the crowding effect induces some smooth biases which increase when the CLF decreases.

Thereafter, to avoid any problem of crowding, the $CLF$ equals to the width of the image for all the following simulations in order to respect the sine-qua-none condition: $D_*<CLF$.\\

\begin{figure}[!h] 
\begin{center}
\begin{tabular}{cccc}
\includegraphics[width =39mm]{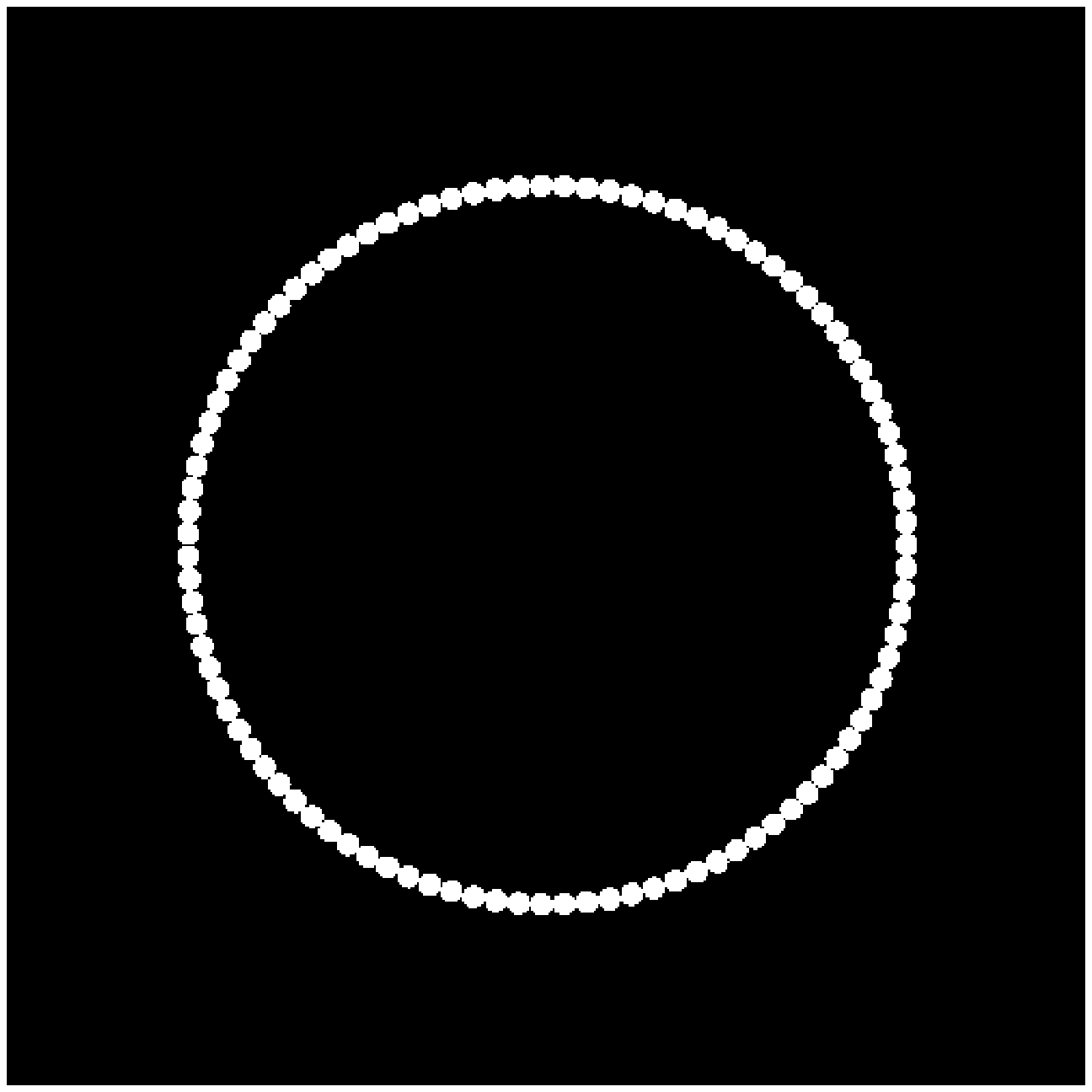} &
\includegraphics[width =39mm]{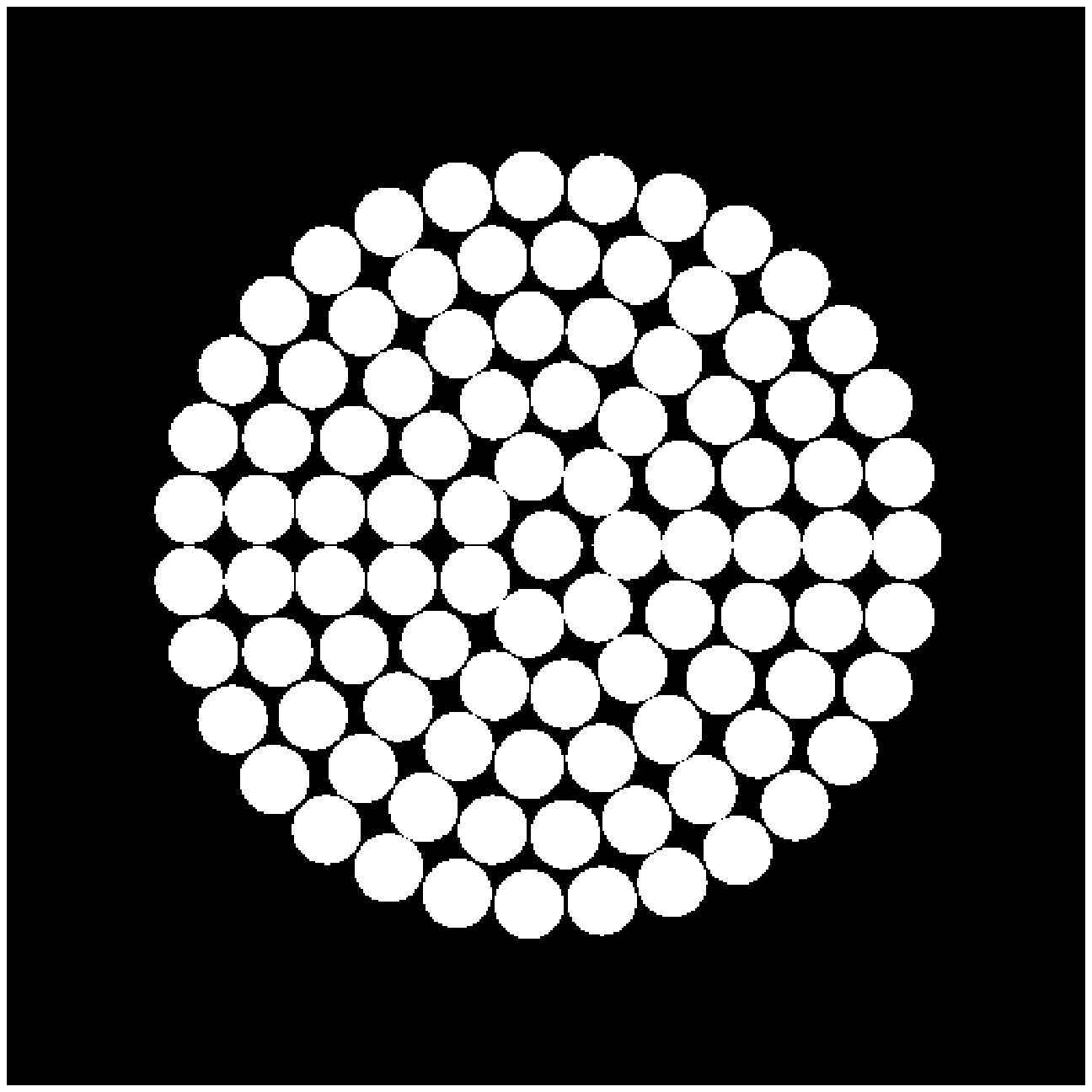} &
\includegraphics[width =39mm]{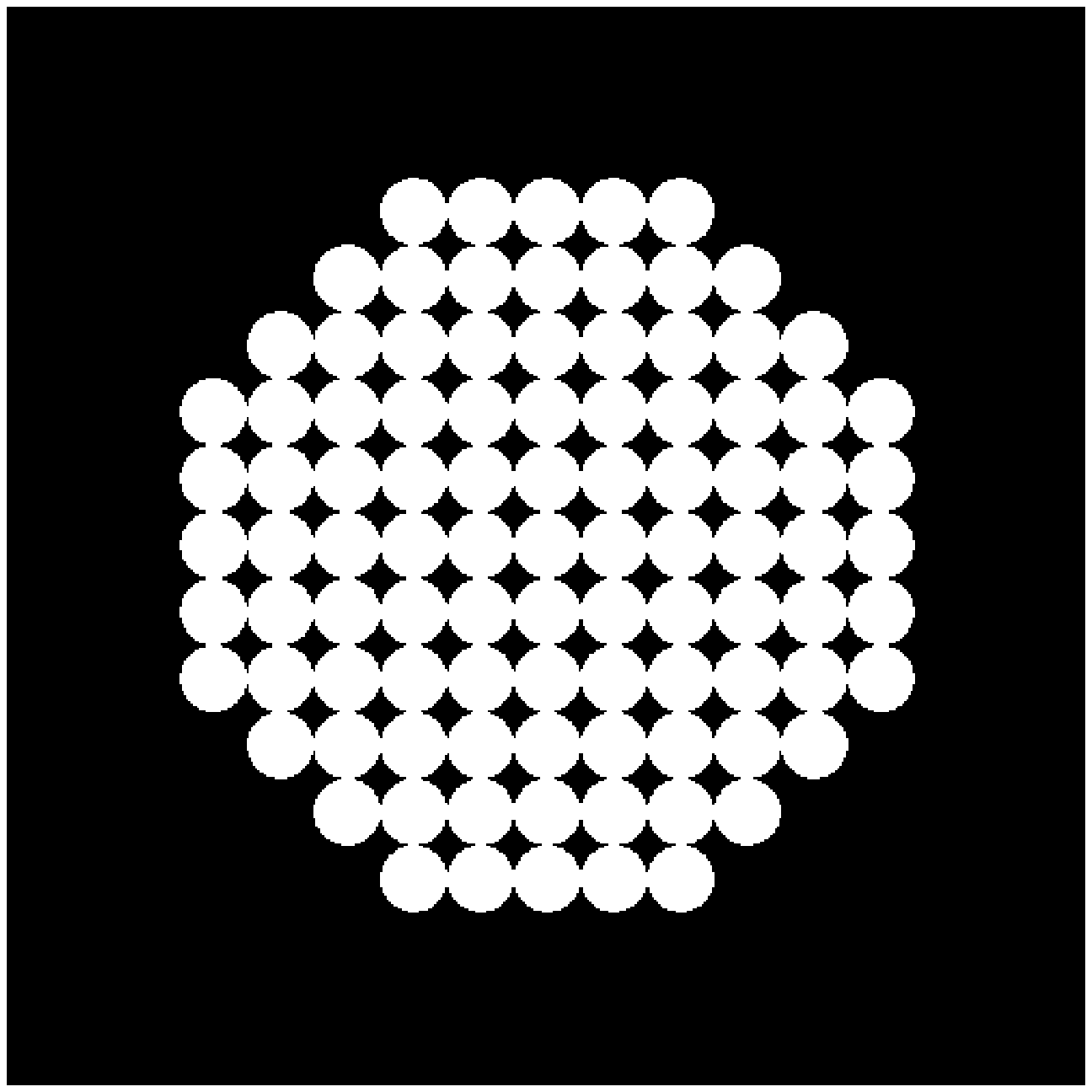} &
\includegraphics[width =39mm]{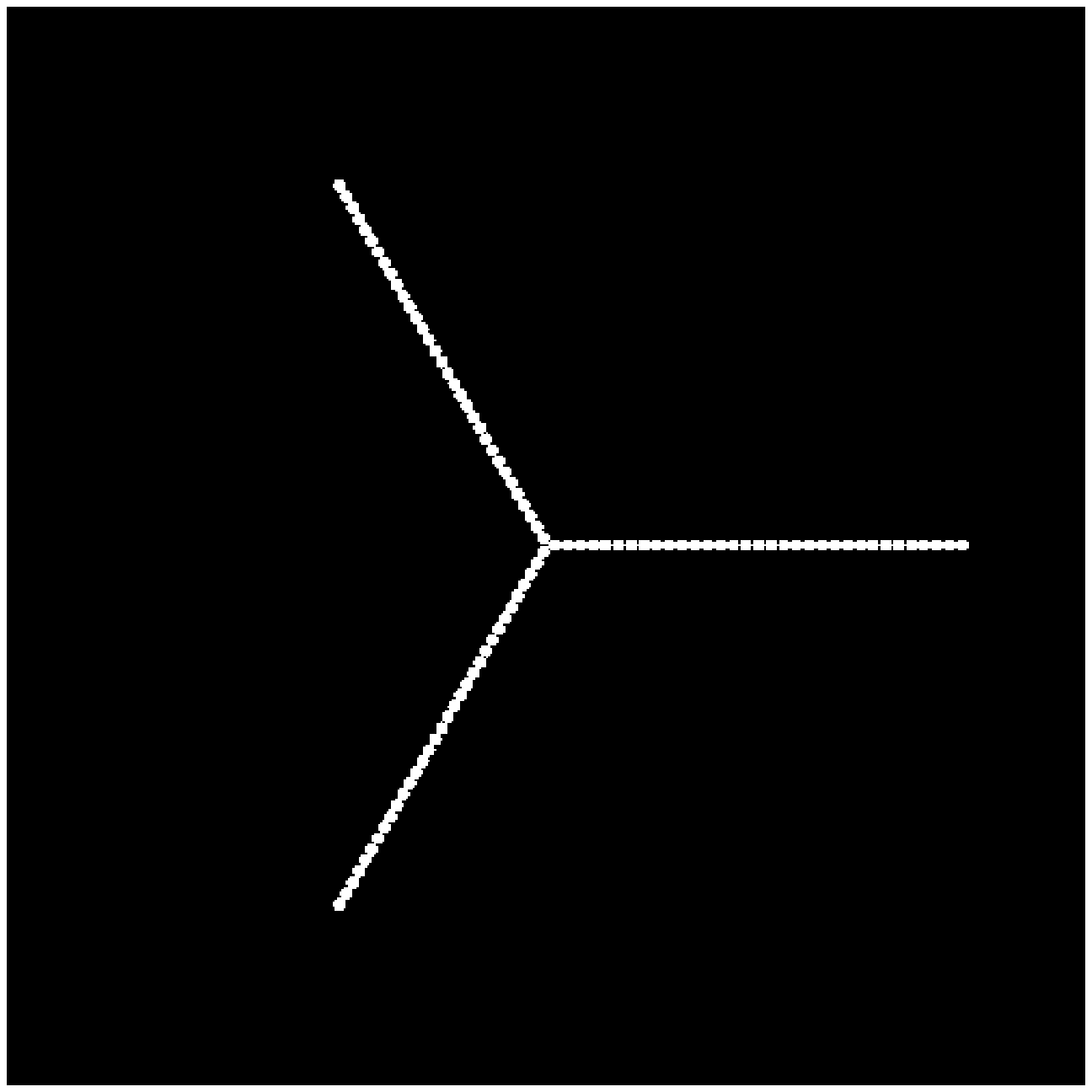} \\
\includegraphics[width=39mm]{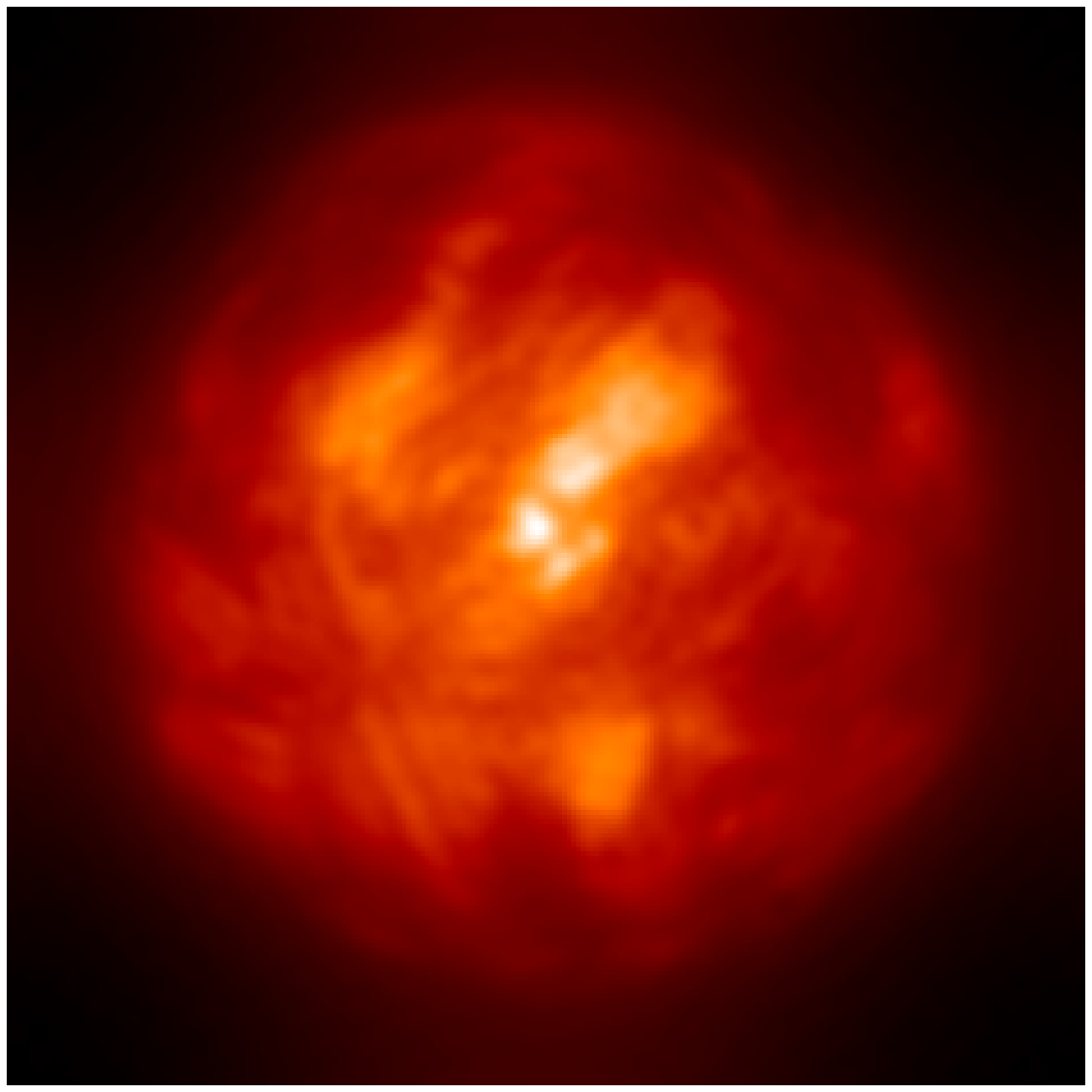} &
\includegraphics[width=39mm]{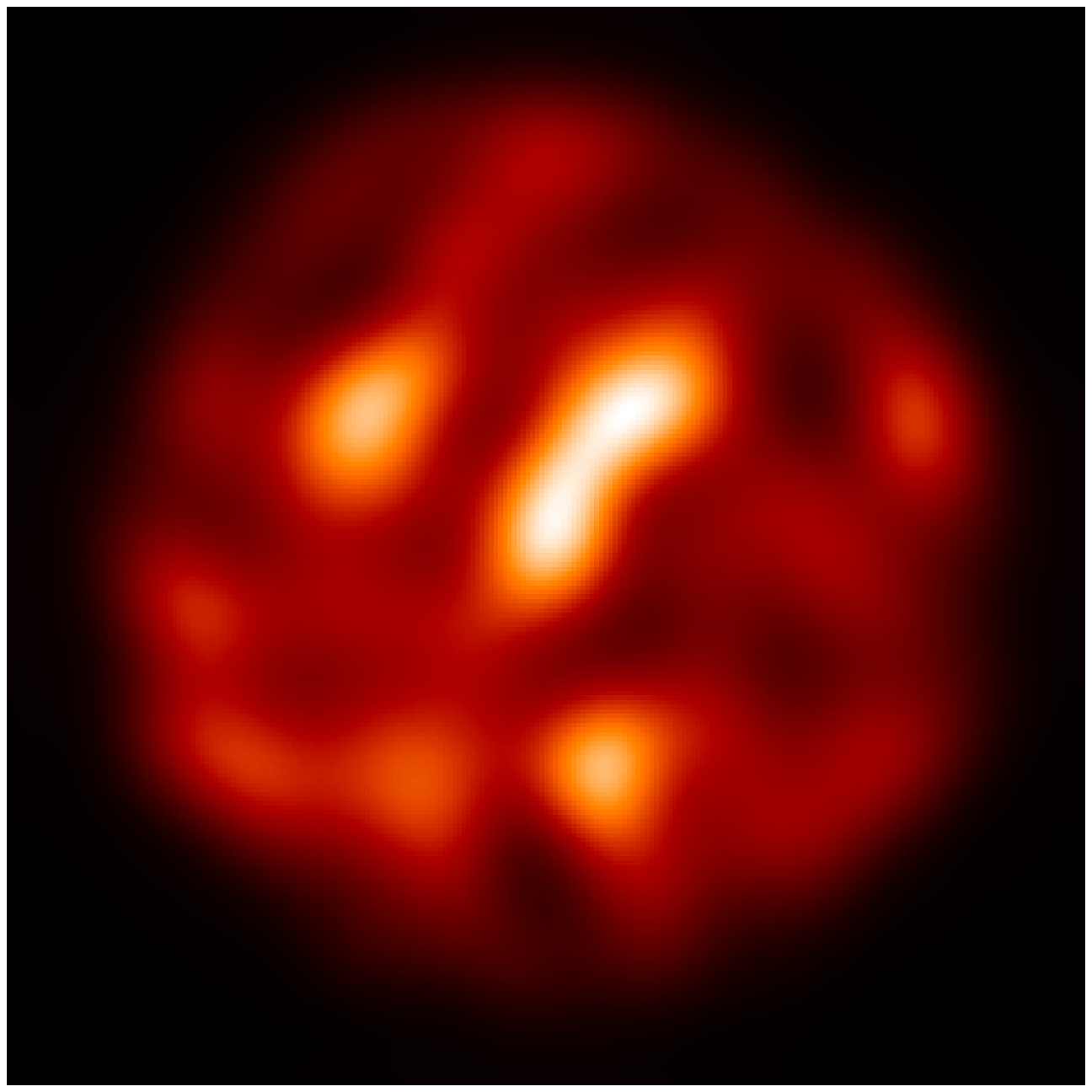} &
\includegraphics[width=39mm]{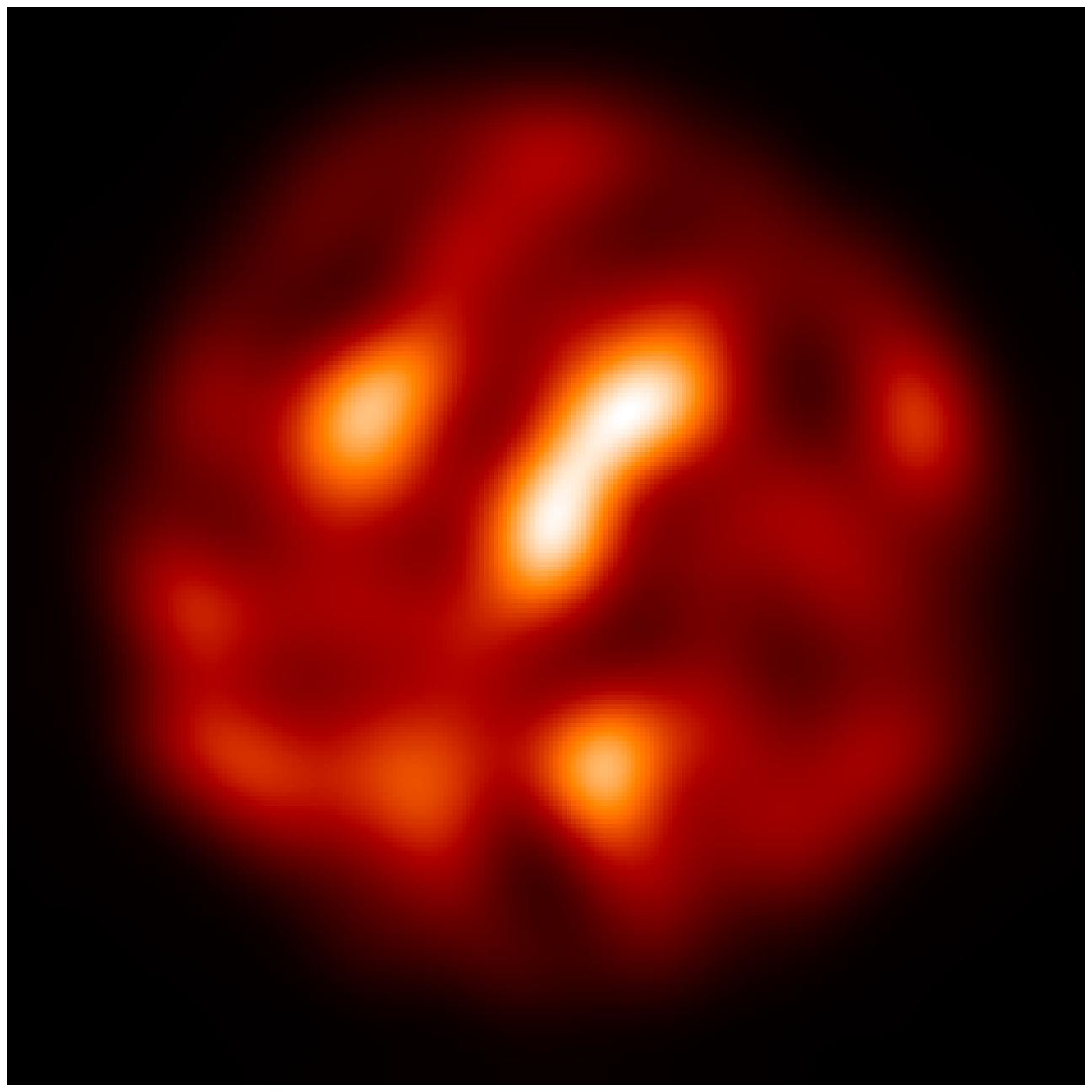} &
\includegraphics[width=39mm]{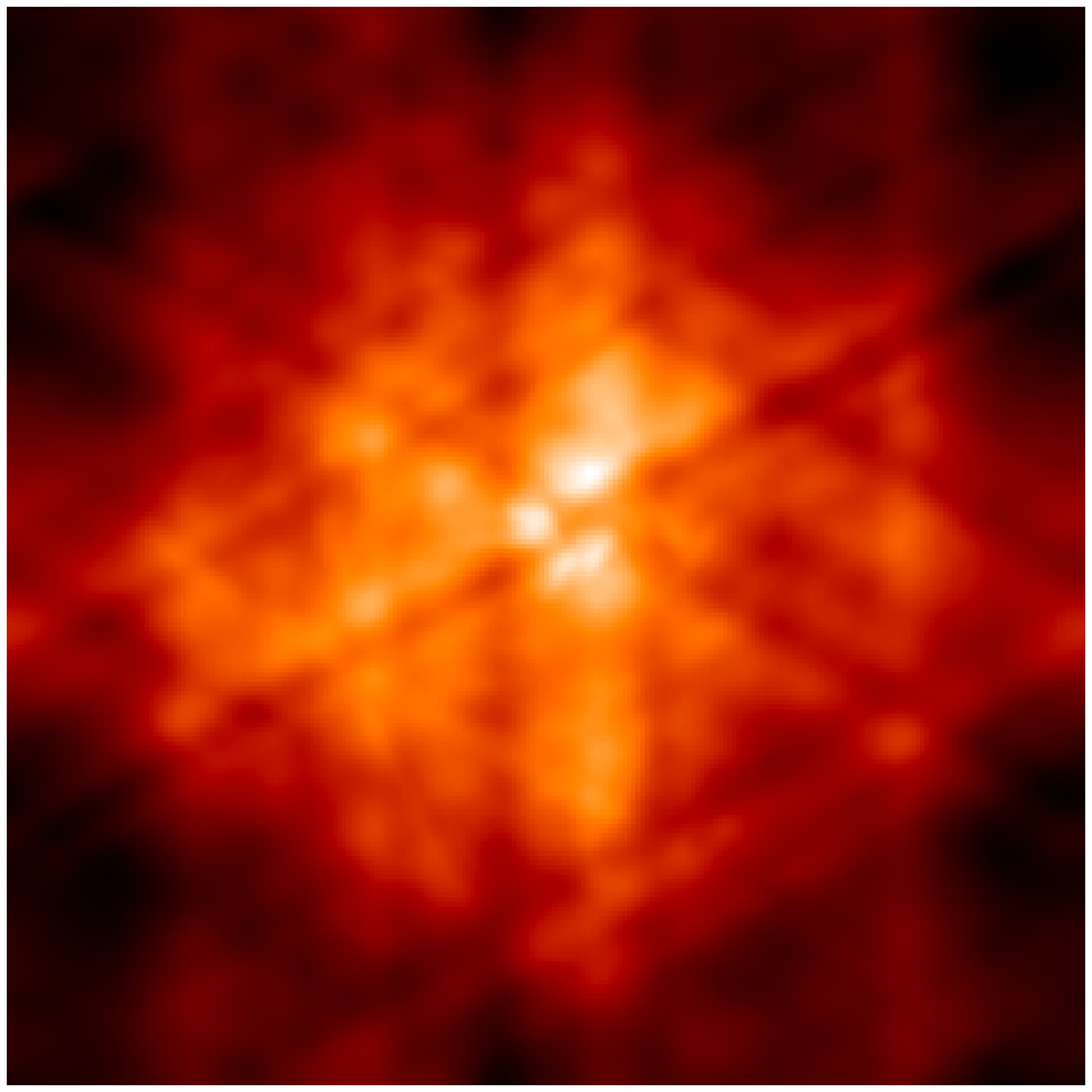} \\
\end{tabular}
\end{center}
\caption[] 
{Direct images of Betelgeuse (bottom) as a function of different array configurations of 100 telescopes having the same Clean FOV of $50~mas$. Here is shown the output densified pupil of each array (top, from left to right): annular configuration ({\it OVLA}), concentric multi-rings configuration ({\it KEOPS}), regular squared configuration ({\it CARLINA}), "Y" configuration ({\it ELSA}). For practical reasons, we use here the names in italic which corresponds to projects of large interferometric arrays \cite{Lardiere2007} even if they are already obsolete.\\
\\
\label{OVLA-ELSA-CARLINA-KEOPS} }
\end{figure} 

\subsection{Influence of the array configuration on the interference function} \label{}

In a general sense for interferometry, the object on the sky is filtered spatially by the array of telescopes, so that the geometry of the array pattern and the number of sub-apertures determine the high angular resolution information sampled on the sky. That is why the choice of the array configuration is of a great importance, and depends on the scientific target.

In the case of a hypertelescope, the image is biased by the halo of the interference function \cite{Lardiere2007}, which induces a contrast loss in the image. The quality of the interference function of the array is simply related to the actual shape of the entrance pupil. If the input sub-pupils are distributed regularly, the densified pupil is almost complete, and the halo inside the Clean FOV reproduces the diffraction pattern of a large monolithic telescope covering all the sub-pupils. 
If the densified pupil shows gaps, additional diffraction figures are added to first ones.
Thus, the halo is minimized by maximizing the densified pupil filling rate $r_o$, with a regular pattern of the sub-apertures in the entrance pupil \cite{Patru2009}.

Figure \ref{OVLA-ELSA-CARLINA-KEOPS} shows the case of 4 different geometries of arrays having the same number of telescopes ($N_T \approx 100$ here) and the same Clean FOV, i.e. the same minimum baseline.
It appears clearly that for a given Clean FOV and a given number of telescopes, the OVLA configuration overcomes the other configurations, since we can see not only the large scale granules but also the inter-granular lanes. Indeed, the spatial distribution of the structures is barely disturbed, but the photometric distribution is biased by the halo of the PSF \cite{Patru2009}. ELSA is clearly the worst one due to its non centro-symetric geometry. The images of CARLINA and KEOPS provide a better contrast between the bright and dark spots, but only the largest structures appear, which is mainly due to the fact that the maximum baseline is 3 times lower than for OVLA so as to keep the same minimum baseline. Note that as long as the object remains inside the Clean FOV and with a significant number of telescopes, the redundancy of the array has no influence (CARLINA or KEOPS provides the same image).

\begin{figure}[!h] 
\begin{center}
\begin{tabular}{cccc}
\includegraphics[width=39mm]{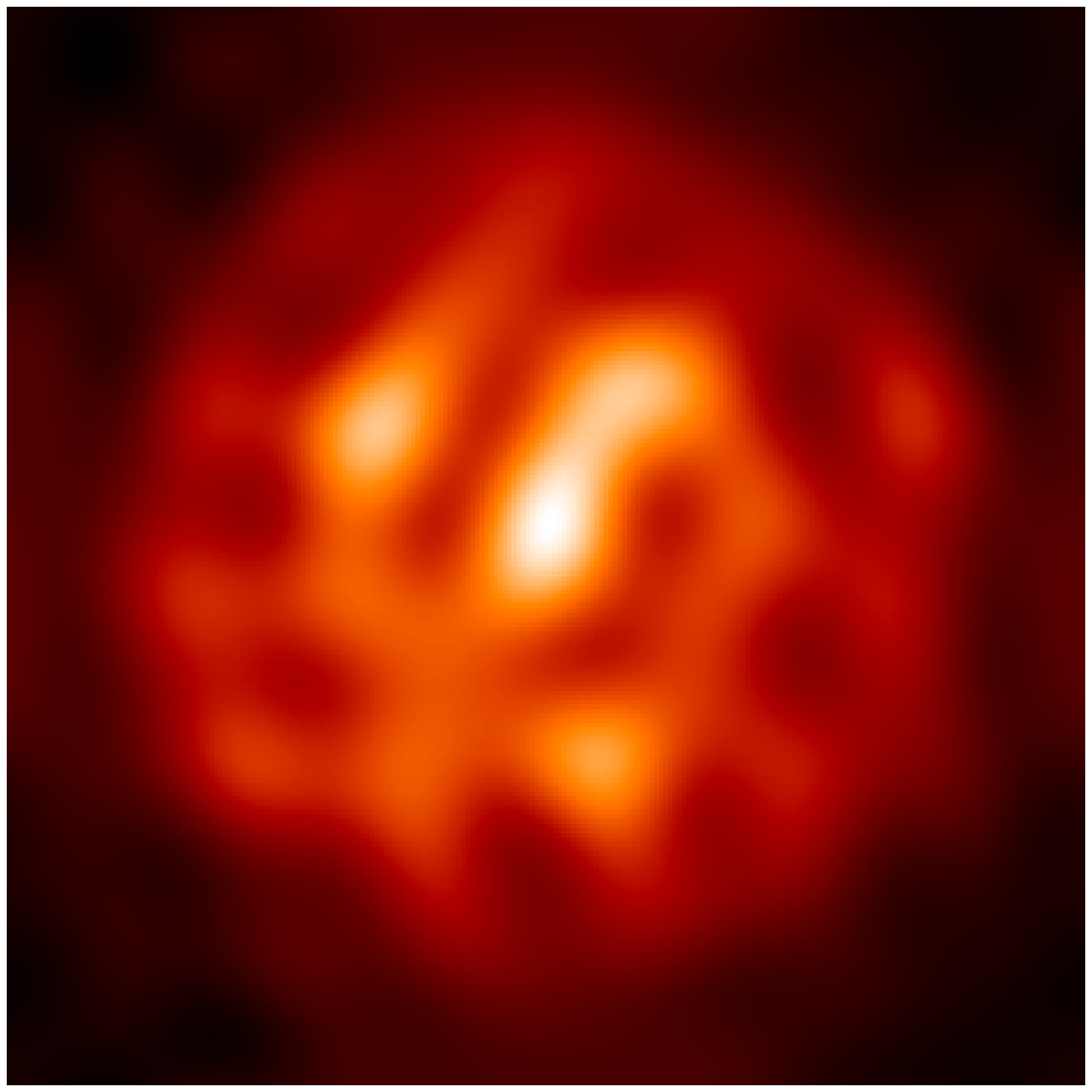} &
\includegraphics[width=39mm]{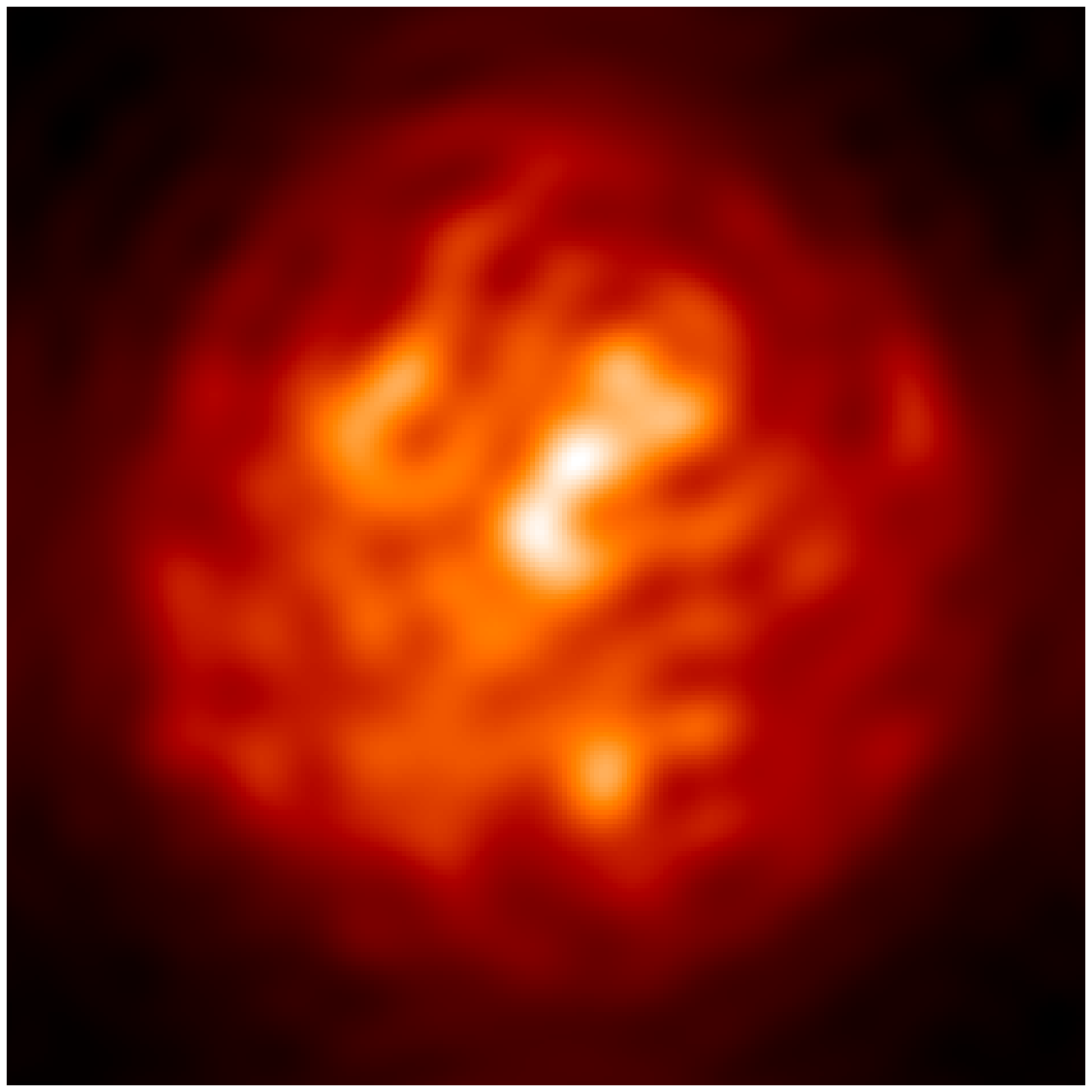} &
\includegraphics[width=39mm]{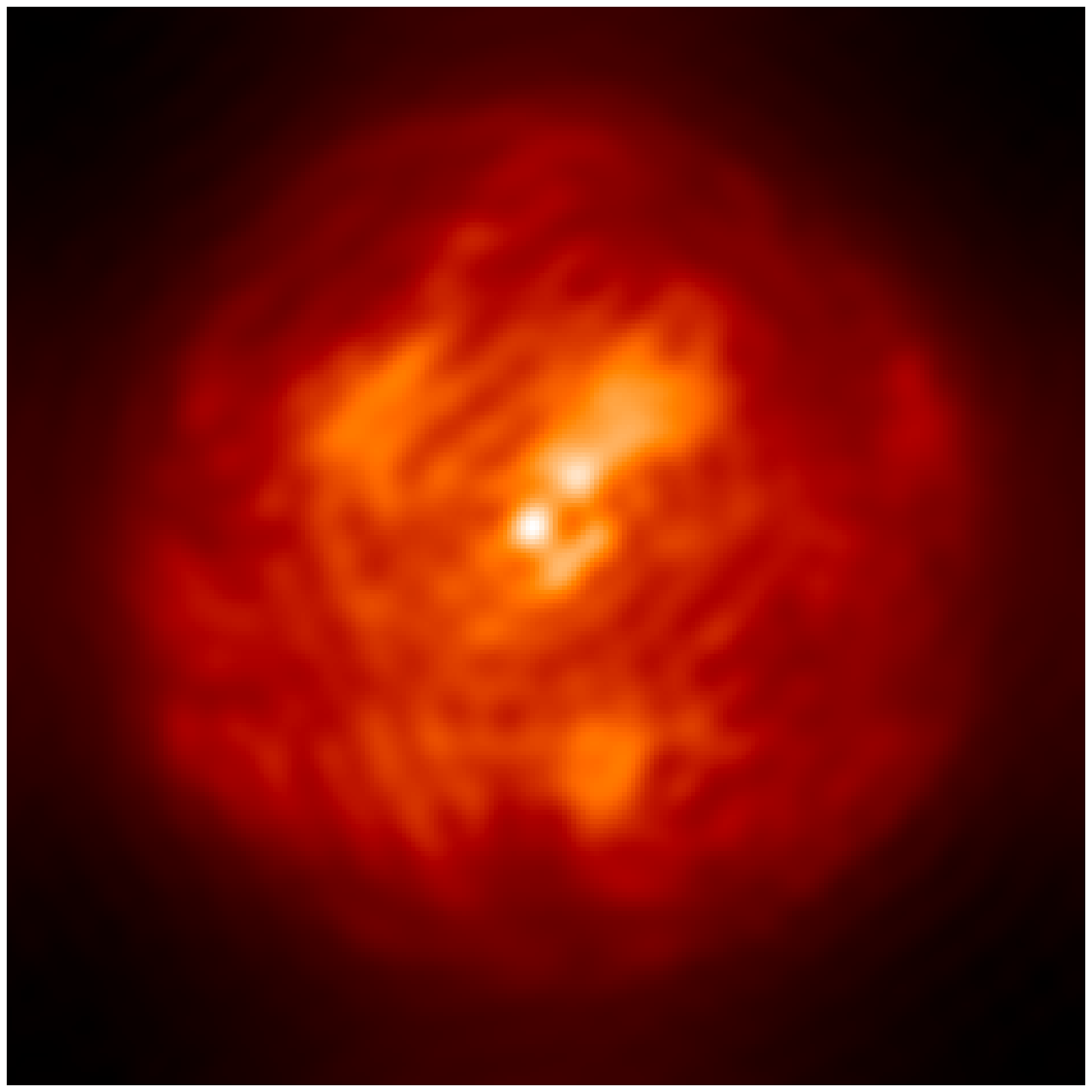} &
\includegraphics[width=39mm]{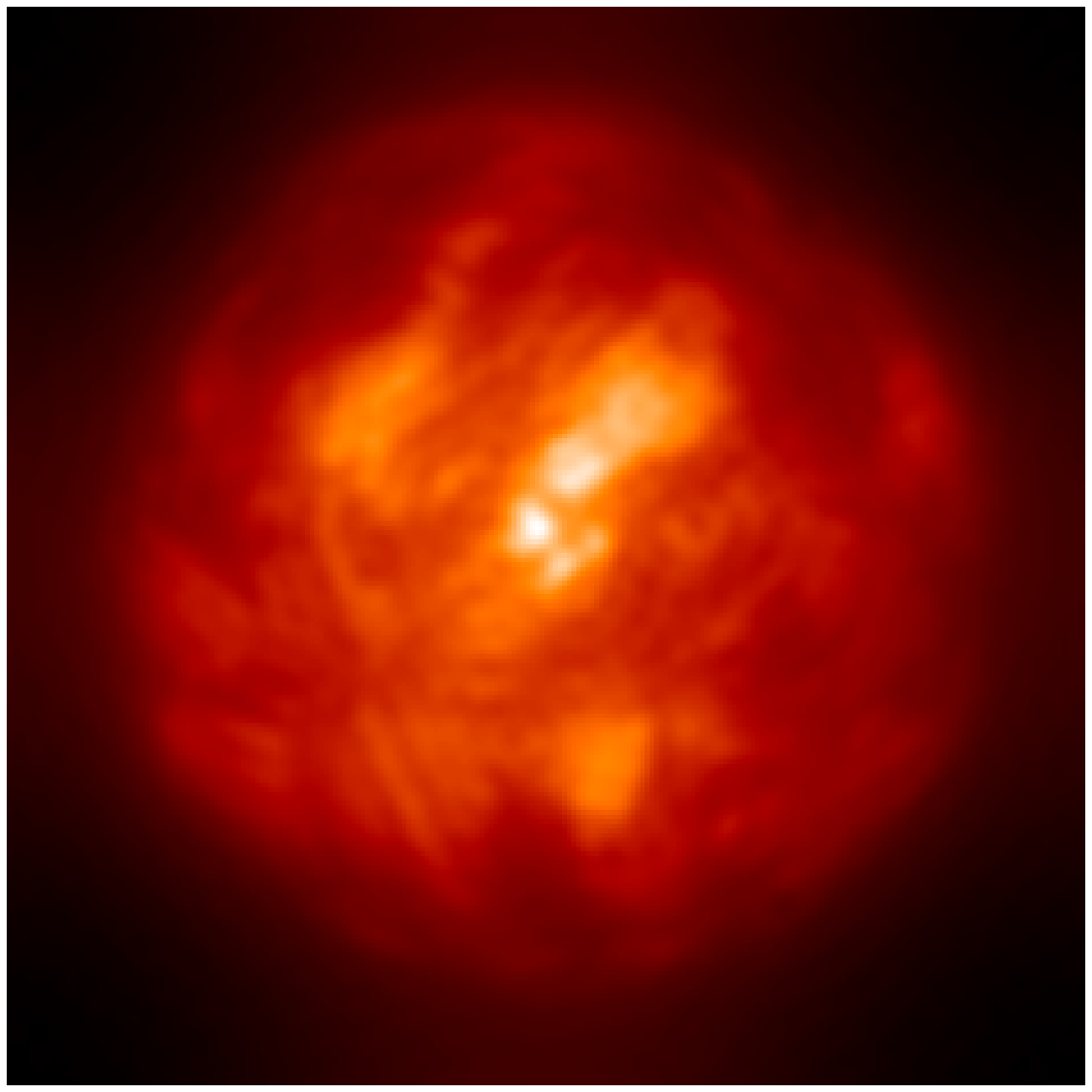} \\
\includegraphics[width=39mm]{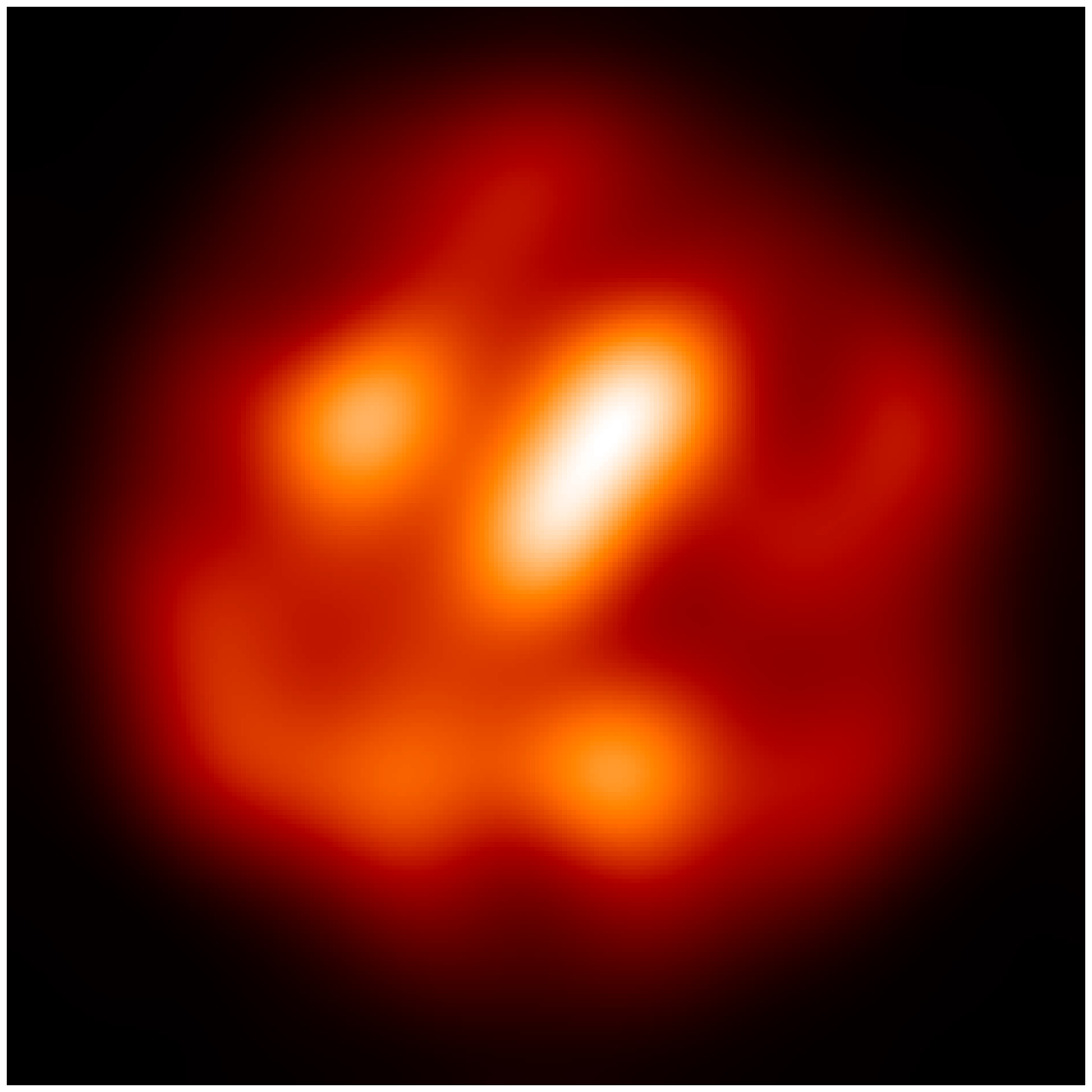} &
\includegraphics[width=39mm]{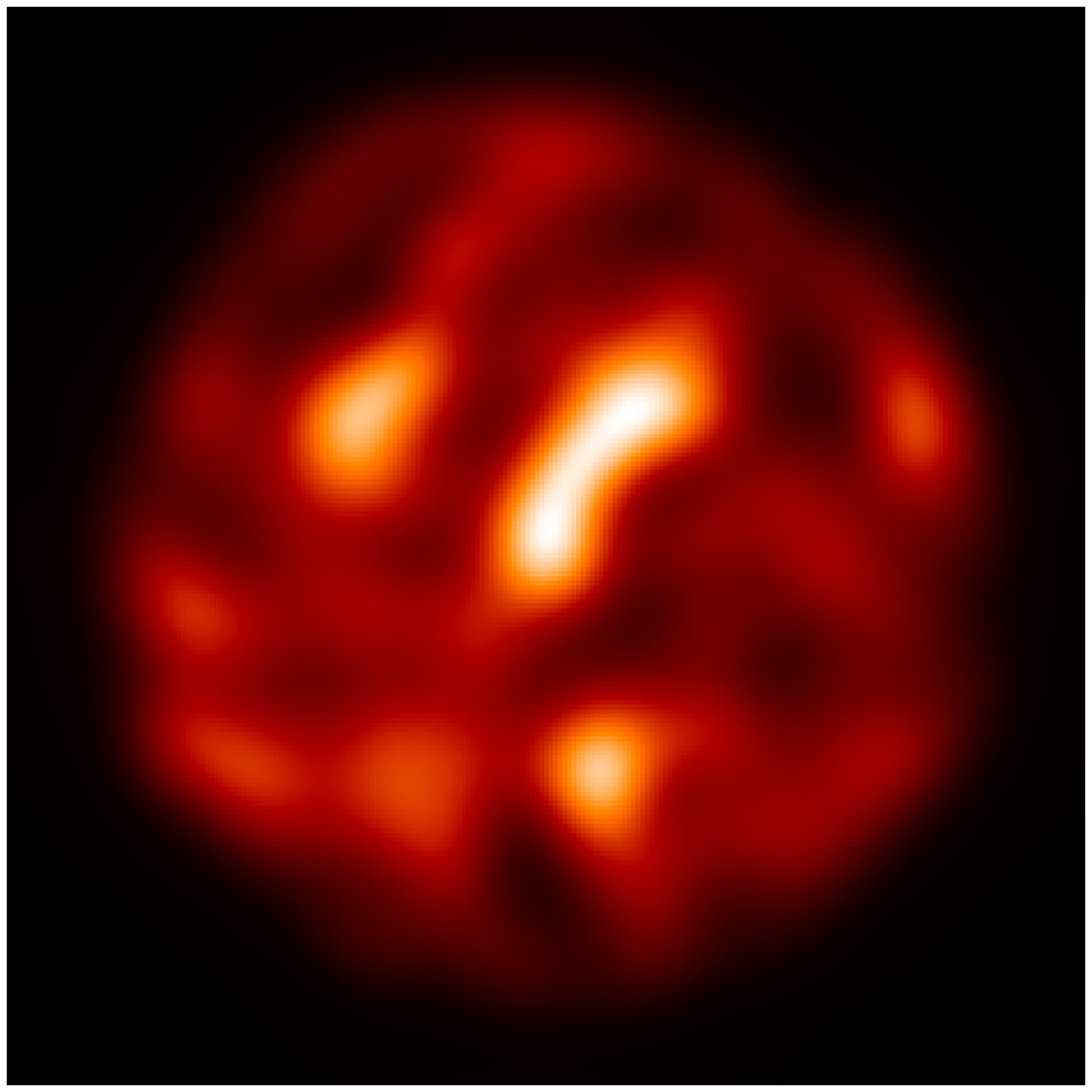} &
\includegraphics[width=39mm]{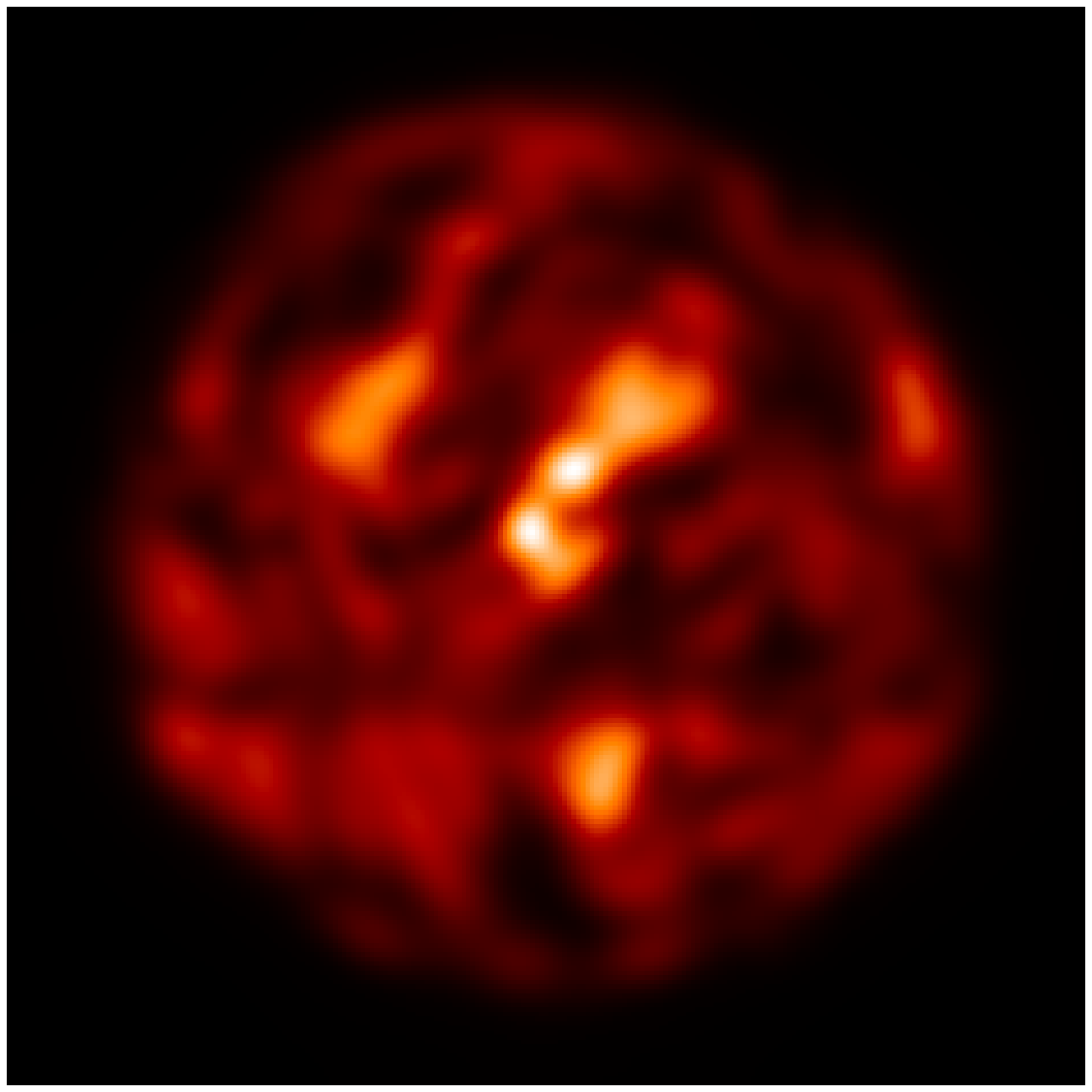} &
\includegraphics[width=39mm]{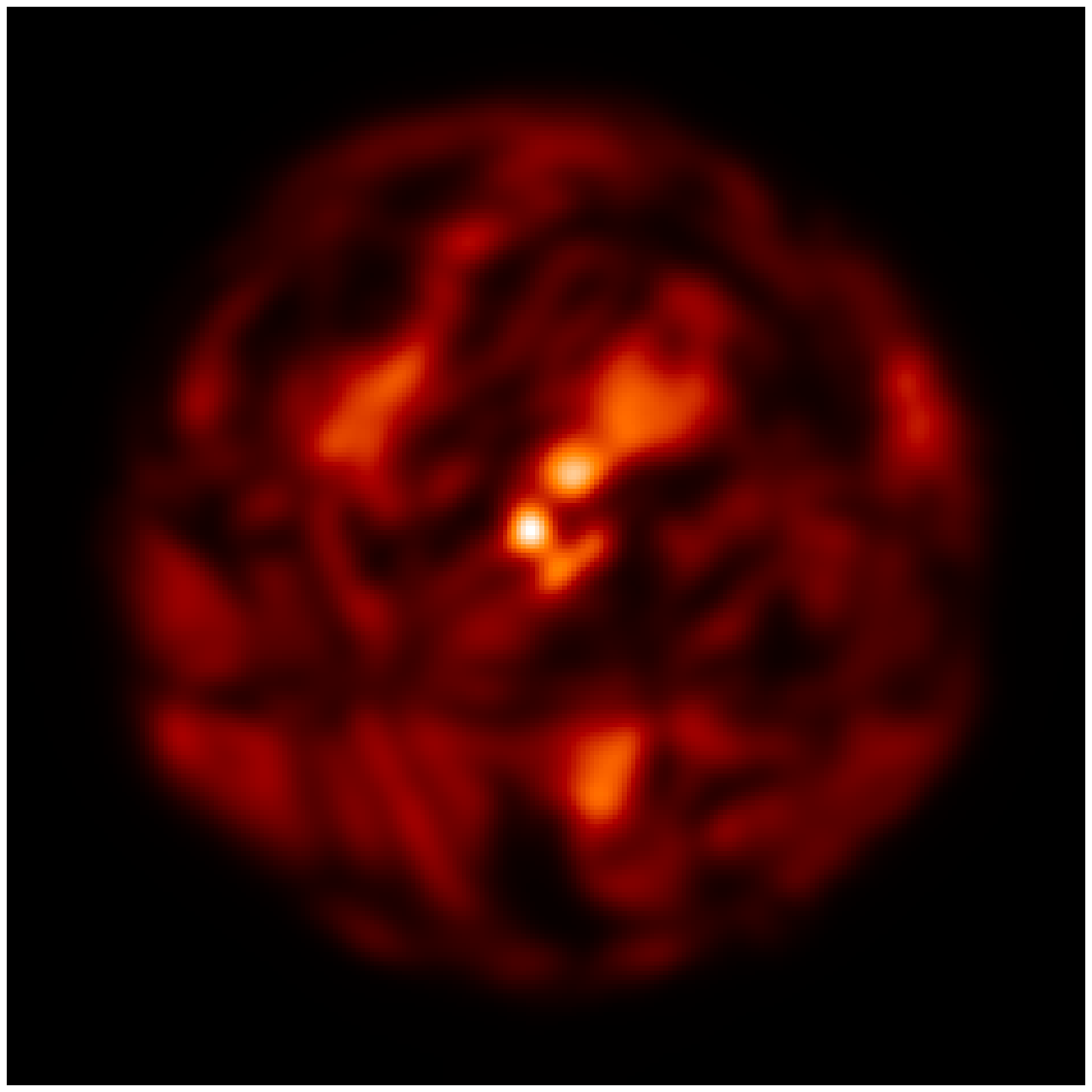} \\
\end{tabular}
\end{center}
\caption[] 
{Direct images of Betelgeuse with the OVLA configuration (top) and the KEOPS configuration (bottom) having the same Clean FOV and an increasing number of telescopes. OVLA20 and KEOPS40 (left) have the same resolution ($\approx 10~mas ~@~745~nm$). In the same way, OVLA36 and KEOPS133 (middle left), OVLA63 and KEOPS408 (middle right), OVLA90 and KEOPS833 (right) have the same resolution (respectively $\approx 5~mas$, $\approx 3~mas$, $\approx 2~mas$  $~@~ 745~nm$).\\
\\
\label{keops_ovla_same_resolution} }
\end{figure}

In an other hand, the figure \ref{keops_ovla_same_resolution} shows the case of the OVLA and KEOPS configurations having the same Clean FOV and the same resolution, i.e. the same minimum and maximum baselines. This is made possible by adjusting the number of telescopes. It appears clearly now that KEOPS is the best one, thanks to its complete densified pupil which provides the best PSF quality with the lowest halo level and the best image with the highest contrast \cite{Patru2009}.
However, KEOPS needs a huge amount of telescopes compared to OVLA. 

\begin{figure}[!h] 
\begin{center}
\begin{tabular}{cccc}
\includegraphics[height=50mm]{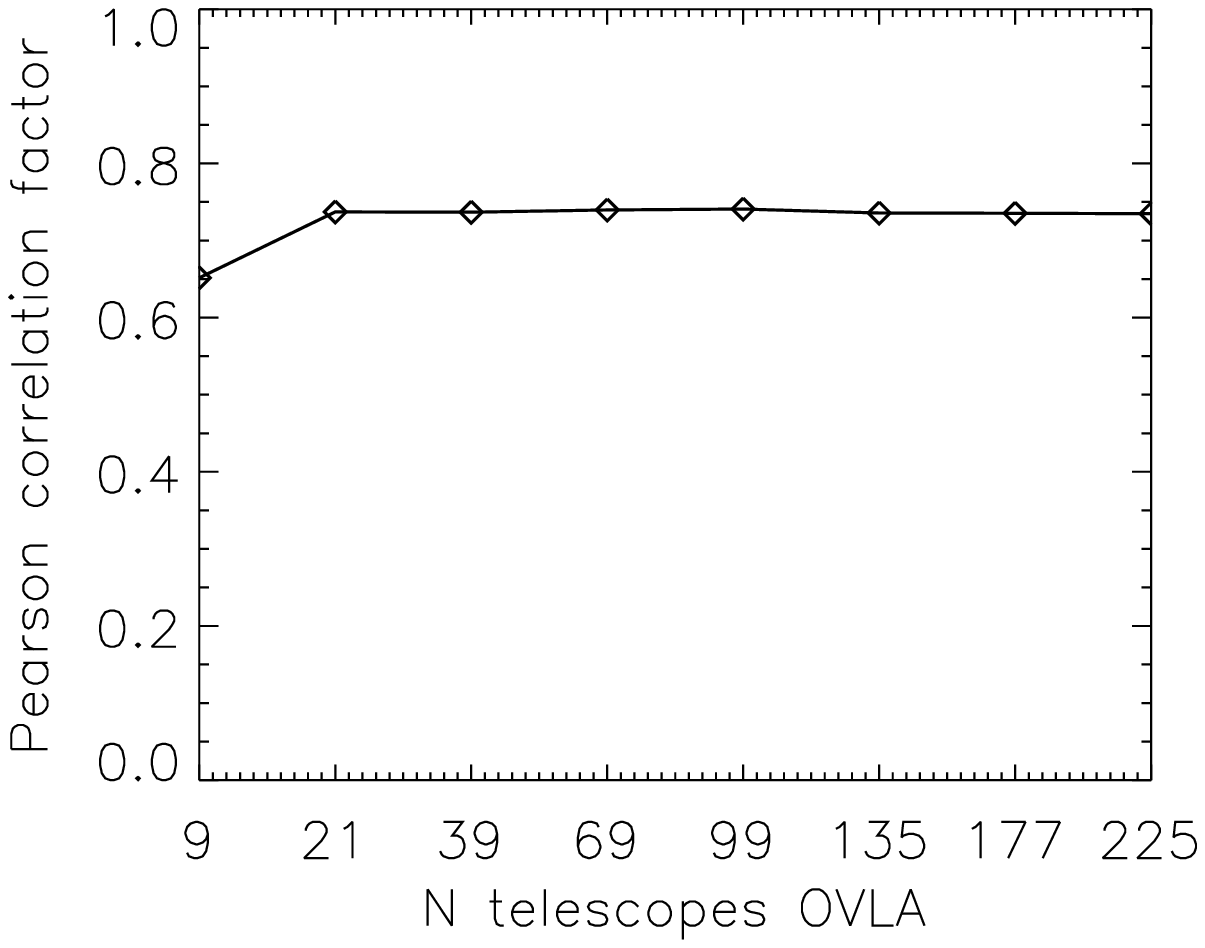} &
\includegraphics[height=50mm]{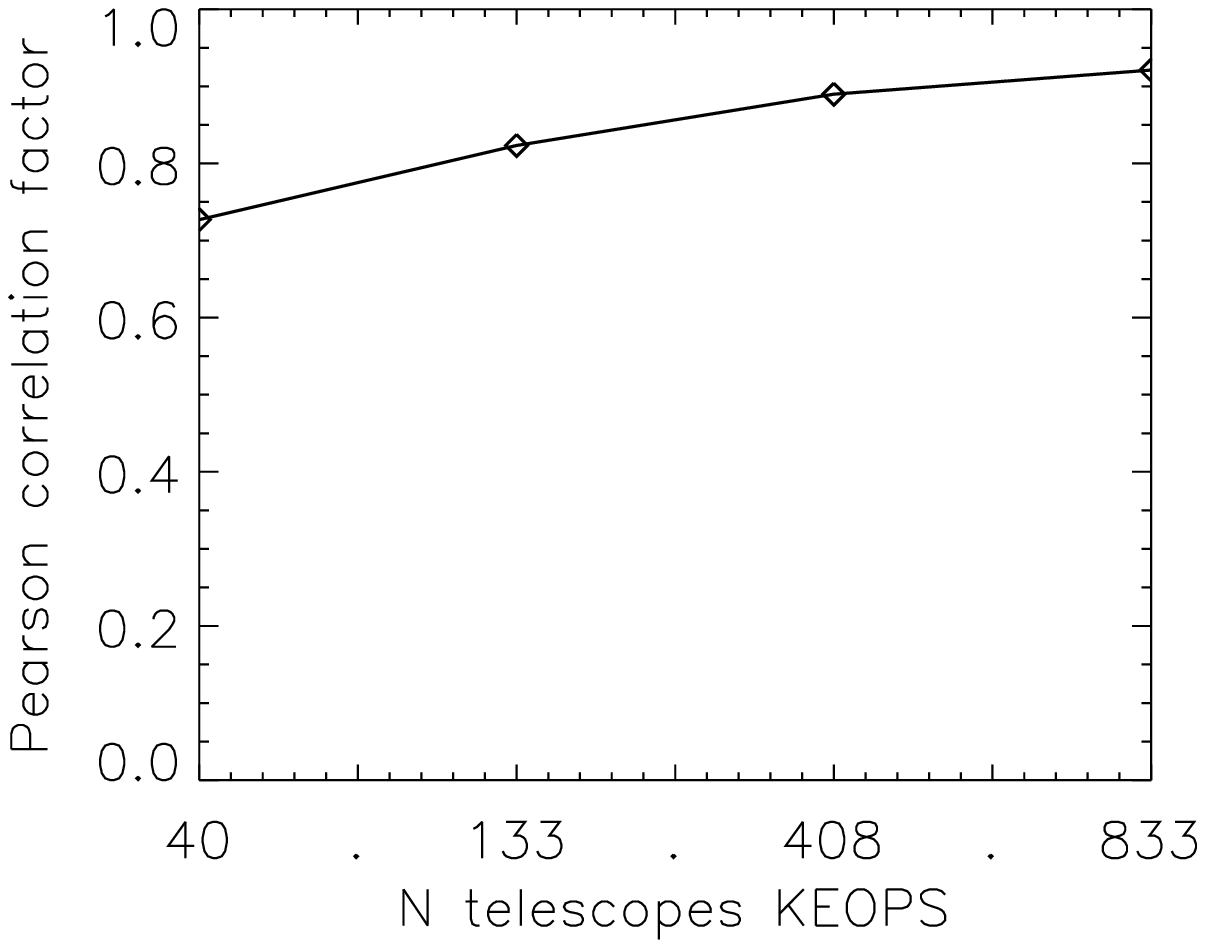} \\
\end{tabular}
\end{center}
\caption[] 
{Pearson correlation factor as a function of the number of telescopes (left with OVLA and right with KEOPS).\\
\\
\label{var_correl_config} }
\end{figure}

Figure \ref{var_correl_config} shows the Pearson correlation factor as a function of the array configuration (geometry and number of telescopes). There is no significant gain of the image quality over 100 telescopes with OVLA. For KEOPS, the image quality increases with the number of telescopes along with the ultimate resolution. Firstly, both large and small scale structures are revealed. Secondly, the dynamic range improves so that the contrast between bright and dark structures tends to the same contrast in the original image (see Fig. \ref{originalBetelgeuse}).\\
\\

\subsection{Influence of the recombination mode on the diffraction envelope} \label{}

All the previous considerations (such as the resolution, the field of view, the crowding effect) are only related to the array, whatever the recombination scheme. We now consider the aspects related to the densified pupil combiner, for which different optical schemes are conceivable \cite{Patru2007}, such as a fibered concept \cite{Patru2008}. The standard scheme implies that the image should be multiplied by the diffraction envelope of an output densified sub-pupil \cite{Lardiere2007}. The width of the diffraction envelope depends on the densification factor $\gamma$, which is the scaling ratio between the output and the input sub-pupils \cite{Lardiere2007, Patru2007}. This factor can be adjusted between $1<\gamma<\gamma_{max}$, where $\gamma=1$ is the Fizeau mode and $\gamma_{max}=B_{min}/D_i$ is the maximum densification when two output densified sub-pupils become tangent. We define also the Direct Imaging Field of view ($DIF=\lambda/ ((\gamma-1).D_i))$) where the photons are concentrated inside the envelope. Note that most of the light falls within the Clean FOV in the maximum densification case ($DIF \approx CLF$) \cite{Lardiere2007}.

Due to the diffraction envelope, the intensification of the signal decreases from the axis as a Bessel function for circular apertures (or a pseudo-gaussian function in the fibered case \cite{Patru2007}), so that the intensity at the edge of the Clean FOV compared to the intensity on-axis is 40\% lower in the maximal densification case and 10\% lower with $\gamma=\gamma_{max}/2$ \cite{Patru2009}.
Thus, a partial densification restitutes a more uniform photometry on the Clean FOV. In this paper, we neglecte this effect by considering a low densified image or a Fizeau image, since this effect is due to the combiner and is not linked to the array design.\\
\\

\begin{figure}[!h] 
\begin{center}
\begin{tabular}{cccc}
\includegraphics[width=180mm]{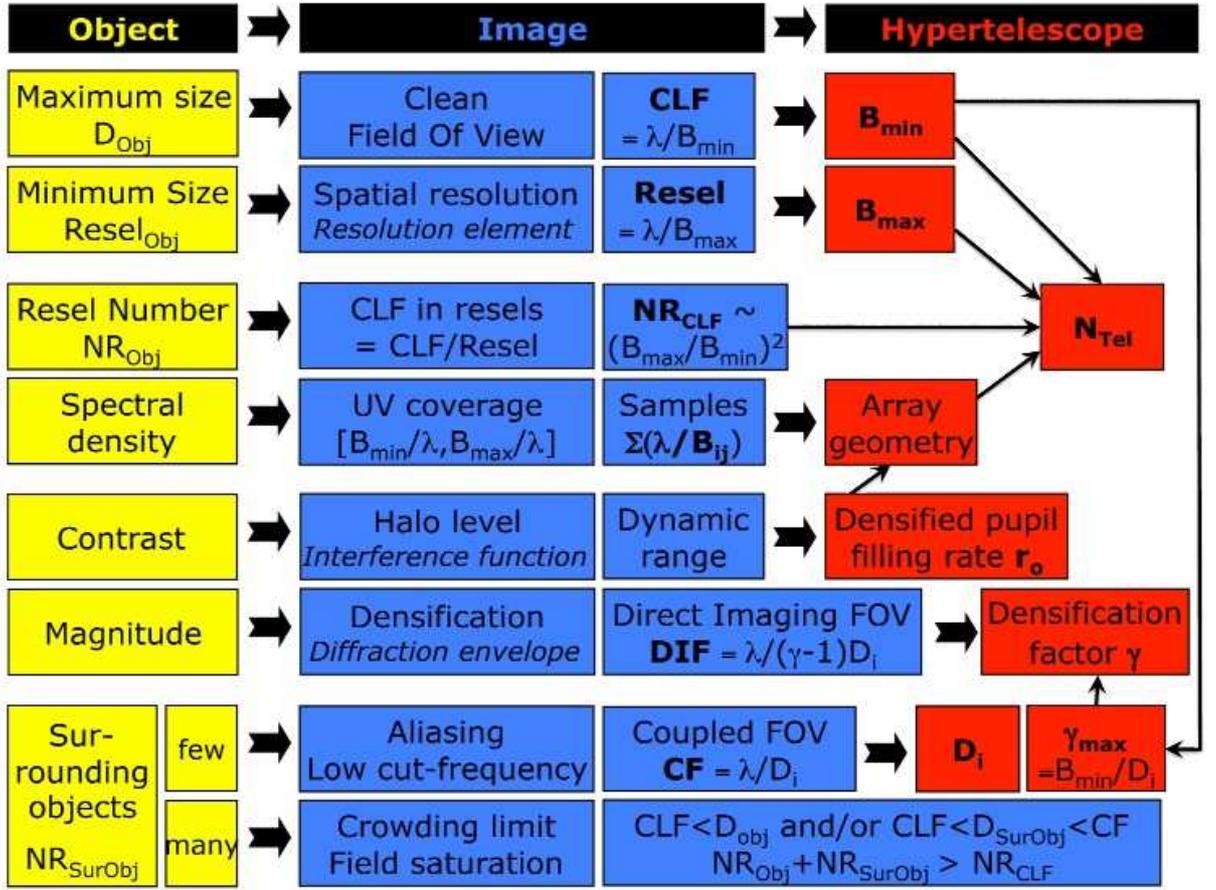}
\end{tabular}
\end{center}
\caption[] 
{Relationship between the astrophysical requirements, the direct image properties and the hypertelescope specifications.\\
\\
\\
\label{htvsobj} }
\end{figure} 

\begin{table}[!h]
\begin{center}       
\begin{tabular}{|l|l|l|l|r|r|r|r|r|} 
\hline
Definition	&	Parameter	&	Formula	&	Unit	&		&		&		&		&		\\
\hline
Central wavelength	&	$\lambda$	&	INPUT	&	mm	&	0.745	&	0.745	&	0.745	&	0.745	&	2	\\
Star diameter	&	$D_*$	&	INPUT	&	mas	&	50	&	50	&	50	&	50	&	50	\\
Image of $N_{Resels}$ x $N_{Resels}$	&	$CLF(Resel)$	&	INPUT	&	$Resel$	&	50	&	25	&	13	&	5	&	5	\\
Resolution element	&	$Resel$	&	$D_*/CLF(Resel)$	&	mas	&	1	&	2	&	4	&	10	&	10	\\
Clean field of view	&	$CLF$	&	 = $D_*$ (HYP)	&	mas	&	50	&	50	&	50	&	50	&	50	\\
Minimum baseline	&	$B_{min}$	&	 $=\lambda/CLF$	&	m	&	3.1	&	3.1	&	3.1	&	3.1	&	8.3	\\
Maximum baseline	&	$B_{max}$	&	 $=\lambda/Resel$	&	m	&	154	&	77	&	40	&	15	&	41	\\
N telescopes for OVLA	&	$N_T$ OVLA	&	 $= 3.8*CLF/Resel$	&		&	190	&	95	&	49	&	19	&	19	\\
N telescopes for KEOPS	&	$N_T$ KEOPS	&	 $\approx 1.5*(CLF/Resel)^2$	&		&	3750	&	938	&	254	&	38	&	38	\\
Maximum densification factor	&	$\gamma_{max}$	&	INPUT	&		&	10	&	10	&	10	&	10	&	10	\\
Densification factor	&	$\gamma$	&	INPUT	&		&	5	&	5	&	5	&	5	&	5	\\
Intensification vs Fizeau	&	$Go$	&	 $= \gamma^2$	&		&	25	&	25	&	25	&	25	&	25	\\
Telescope diameter	&	$D_i$	&	 = $B_{min}/\gamma_{max}$	&	m	&	0.31	&	0.31	&	0.31	&	0.31	&	0.83	\\
Coupled field of view	&	$CF$	&	 $=\lambda/ D_i$	&	mas	&	500	&	500	&	500	&	500	&	500	\\
Direct imaging field of view	&	$DIF$	&	 $=\lambda/ ((\gamma-1).D_i))$	&	mas	&	125	&	125	&	125	&	125	&	125	\\
\hline 
\end{tabular}
\end{center}
\caption{Technical parameters of a hypertelescope to image Betelgeuse in a field of view of $50~mas$ for different size of images expressed in number of $Resels$.
\label{tab:param}}
\end{table} 

\subsection{Object-Image-Hypertelescope relationship} \label{}

Figure \ref{htvsobj} summarizes how to define the general design of the hypertelescope as a function of the scientific requirements \cite{Patru2009}.

The main dimensions of the object are the external diameter and the smallest $Resel$ of interest. 
The maximum size of the object should not exceed the diameter of the clean field, which leads to the value of the minimum baseline of the array ($CLF=\lambda/B_{min}$). 
The smallest $Resel$ corresponds to the required resolving power, which imposes the largest baseline of the array ($Resel=\lambda/B_{max}$).

The complexity of the object determines the required number of $Resels$ in the image, which depends on the number of telescopes and on the array geometry. The other aspect to be considered is the spectral density of the object (i.e. its spatial frequencies distribution) in comparison with the (u,v) plan coverage of the array on the range $[B_{max}/\lambda \, ,\, B_{min}/\lambda]$, which provides a set of resolution samples $\lambda/B_i$ in the image for each baseline $B_i$.

The image quality, as well as the dynamic range, is directly related to the array geometry, which can be optimized by maximizing the densified pupil filling rate $r_o$.
The limiting magnitude is directly related to the performances of the cophasing device, allowing long exposures. It depends also on the densification factor in the presence of read-out noise.

To avoid aliasing and crowding, the object should not exceed the Clean FOV.
Moreover, the diameter of the sub-apertures can be enlarged to avoid the aliasing of surrounding stars.
In the case of Betelgeuse, the closest bright star (BD+07 1055B with a magnitude V 14.5 remote of $42~as$) is so far that the aliasing cannot occur (as long as the telescope diameter is higher than $7~mm$ in the visible).
Moreever, the presence of the circumstellar envelope around Betelgeuse can be problematic, but its luminosity remains negligible in the visible compared to the photosphere \cite{Kervella2009}.

Table \ref{tab:param} gives the main parameters of the hypertelescope to image Betelgeuse in a field of view of $50~mas$. For instance, an image of 50x50 $Resels$ with $1~mas$ of resolution at $745~nm$ of wavelength requires an array with $B_{min}=3.1~m$ and $B_{max}=154~m$, 190 telescopes for OVLA and 3750 telescopes for KEOPS. The maximum densification factor is chosen to 10 so that the telescope diameter equals to $D_i=0.31~m$. The coupled FOV is 10 times larger than the Clean FOV ($CF=500~mas$). A half-densification of $\gamma=\gamma_{max}/2=5$ leads to a Direct Imaging FOV equal to $DIF=125~mas$, where the photons are concentrated. Note that the ELT (diameter of $42~m$) equipped with an aperture masking and a pupil densifier can provide a direct image of 13x13 $Resels$ in the visible ($@~745~nm$) and an image of 5x5 $Resels$ in the infrared ($@~2.2~\mu m$).\\
\\

\section{Technical specifications versus conditions of observation} \label{}

\subsection{Effect of the atmospheric pistons and cophasing aspects} \label{}

\begin{figure}[!t] 
\begin{center}
\begin{tabular}{cccc}
\includegraphics[width=39mm]{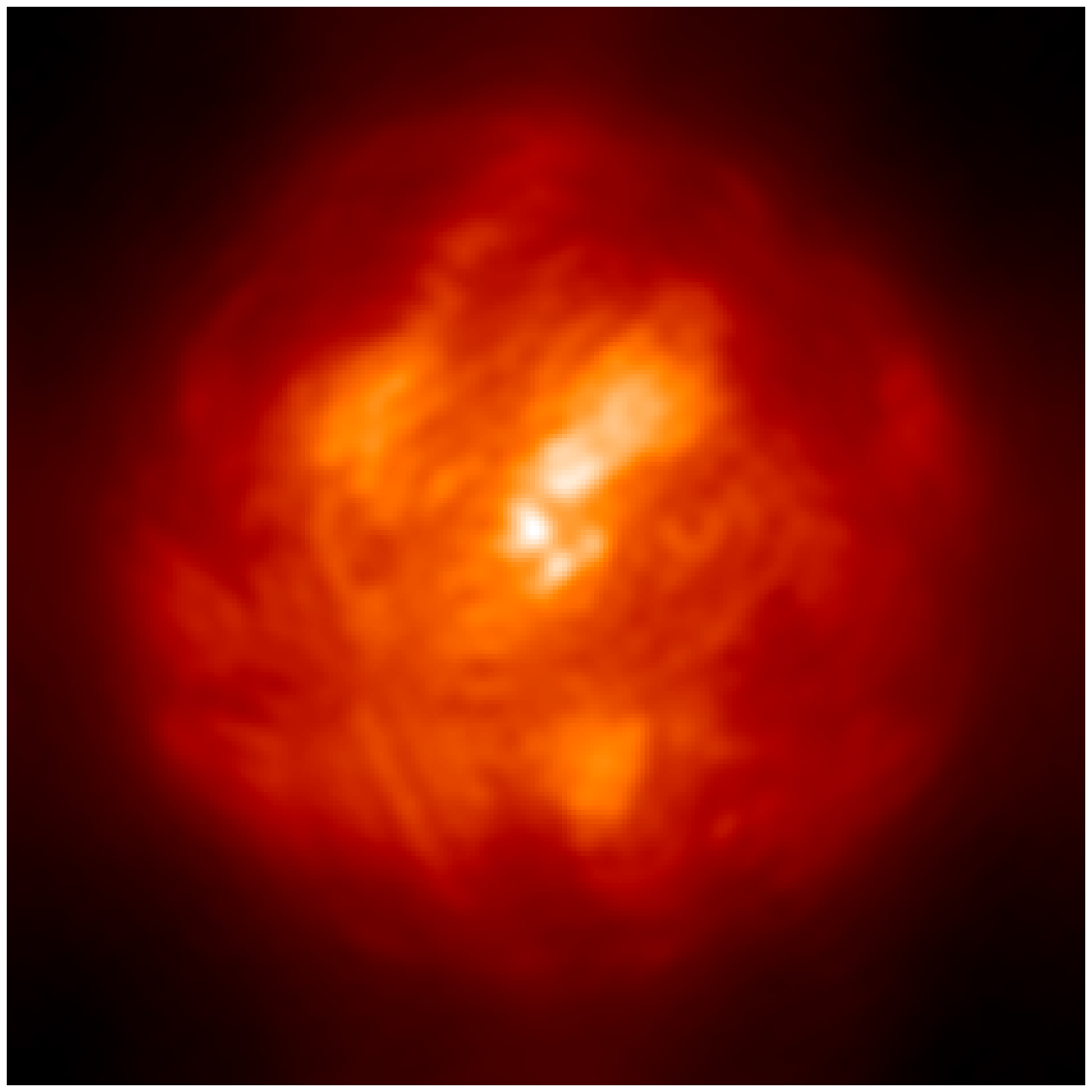} &
\includegraphics[width=39mm]{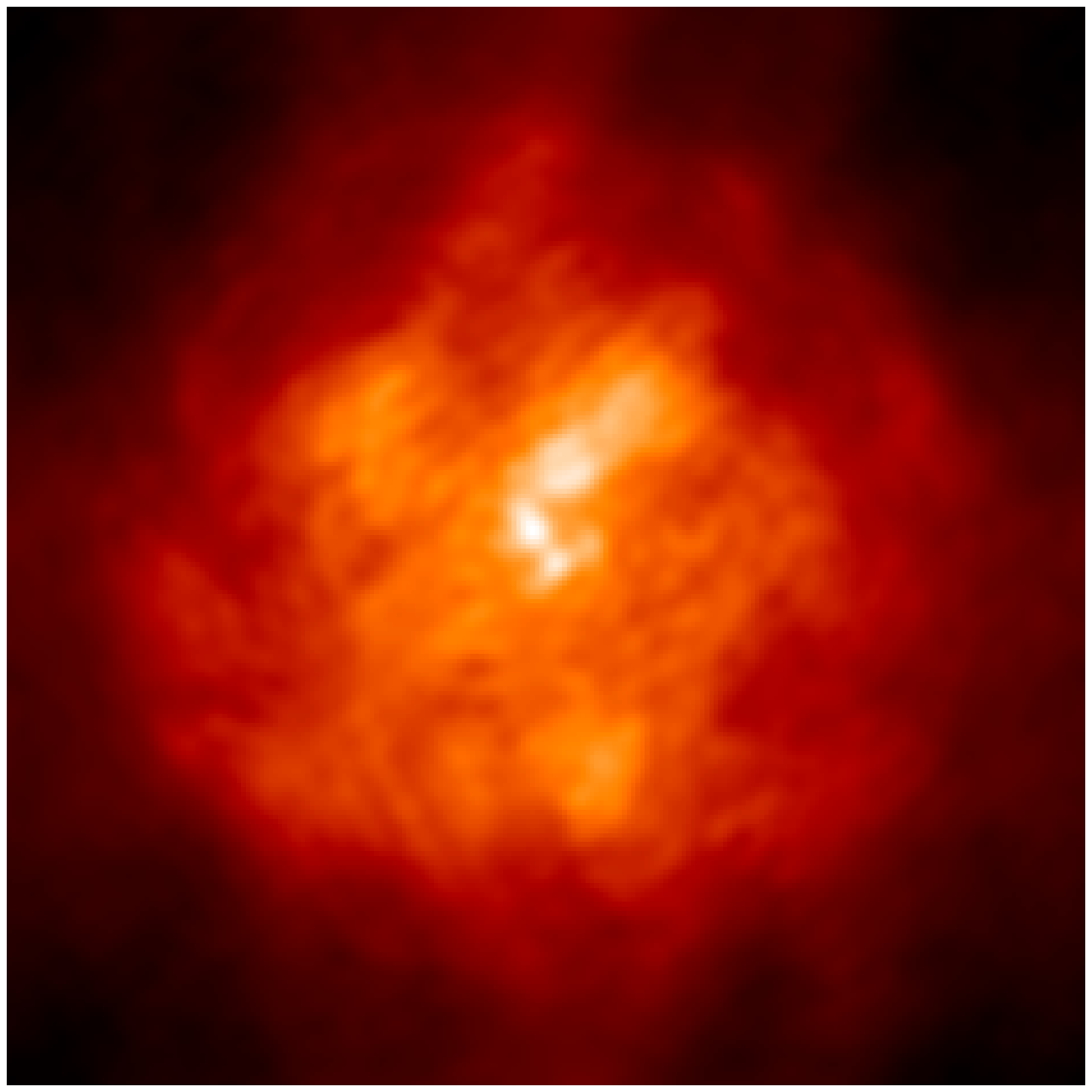} &
\includegraphics[width=39mm]{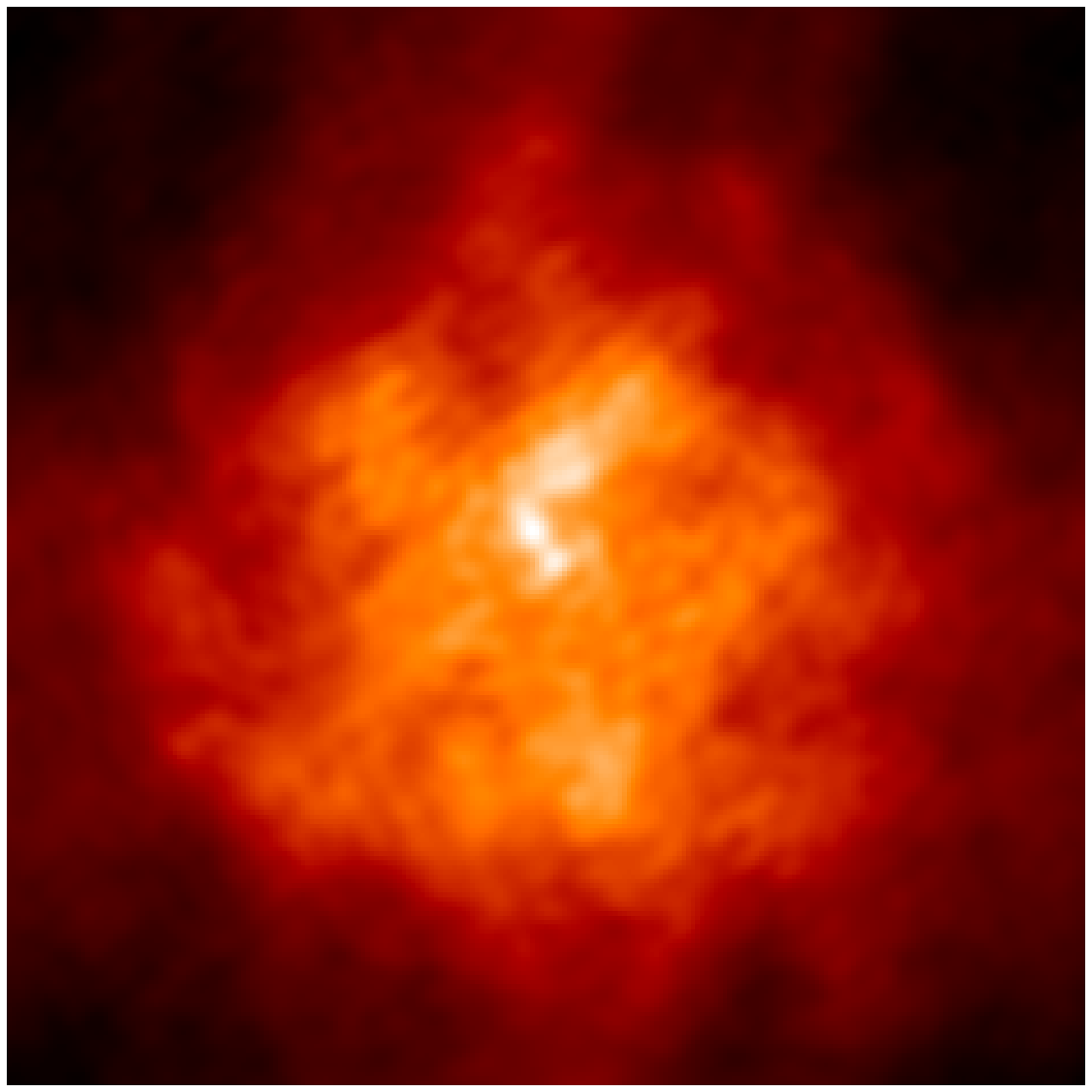} &
\includegraphics[width=39mm]{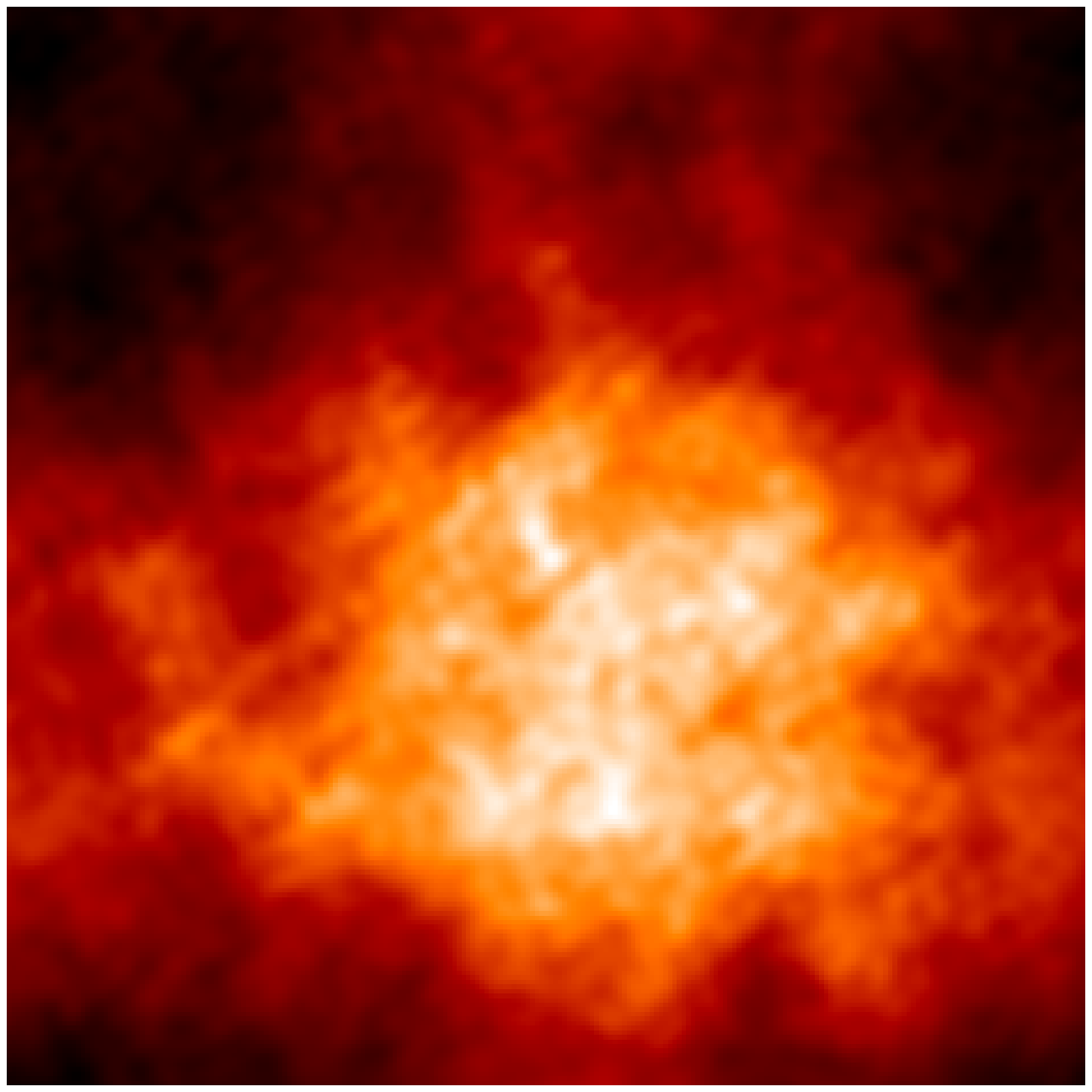} \\
\end{tabular}
\end{center}
\caption[] 
{Direct images of Betelgeuse with the OVLA configuration of 100 telescopes as a function of the residual optical path difference (OPD) between the beams. The standard deviation of the OPD equals to $\lambda/24$,  $\lambda/8$,  $\lambda/6$,  $\lambda/4$ (from left to right).\\
\\
\label{simuOVLA100vsOPD} }
\end{figure}

\begin{figure}[!t] 
\begin{center}
\begin{tabular}{cccc}
\includegraphics[height=60mm]{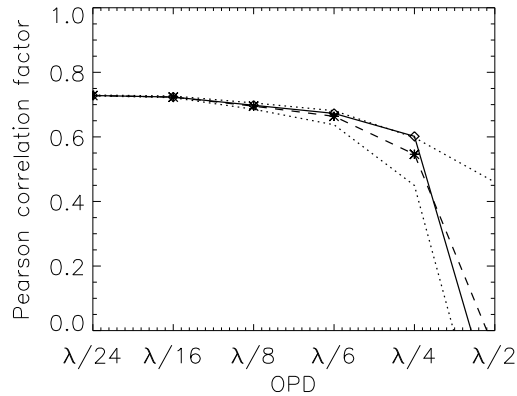}
\end{tabular}
\end{center}
\caption[] 
{Pearson correlation factor as a function of the standard deviation of the optical path difference (OPD) between the beams. The plots correspond to the average value (asterisk + dashed line) surrounded by the minimum and the maximum value (dotted line) of a data set of 100 frames, together with the value of the integrated data set (solid line).\\
\\
\\
\label{var_correl_simuOVLA100vsOPD} }
\end{figure}

Figures \ref{simuOVLA100vsOPD} and \ref{var_correl_simuOVLA100vsOPD} show the effect of the residual optical path difference (OPD) between the beams. As already known \cite{Lardiere2004}, the correction of the atmospheric piston is the hard point. An OPD standard deviation better than OPD $>\lambda/24$ provides the best image quality and a cophasing specification of OPD $>\lambda/8$ is at least required. Otherwise, the small scale structures first and the large scale structures next are destroyed. This specification has also been demonstrated on a fibered testbed \cite{Patru2008}.
Note that the integration of few hundreds frames allows to recover partially the spatial structures in the image if the OPD standard deviation remains higher than OPD $>\lambda/4$. \\

\subsection{Effect of the photometric fluctuations} \label{}

\begin{figure}[!t] 
\begin{center}
\begin{tabular}{cccc}
\includegraphics[width=39mm]{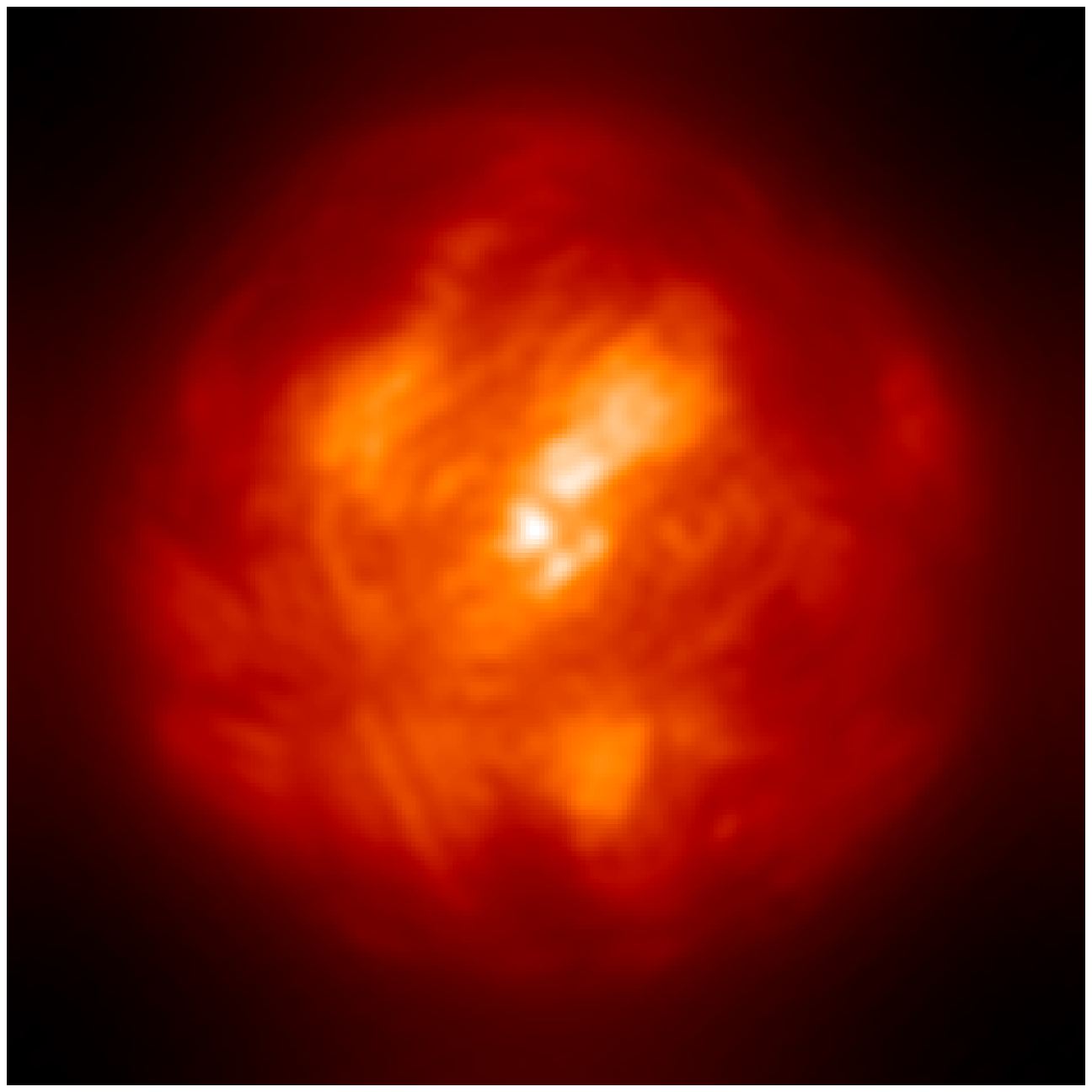} &
\includegraphics[width=39mm]{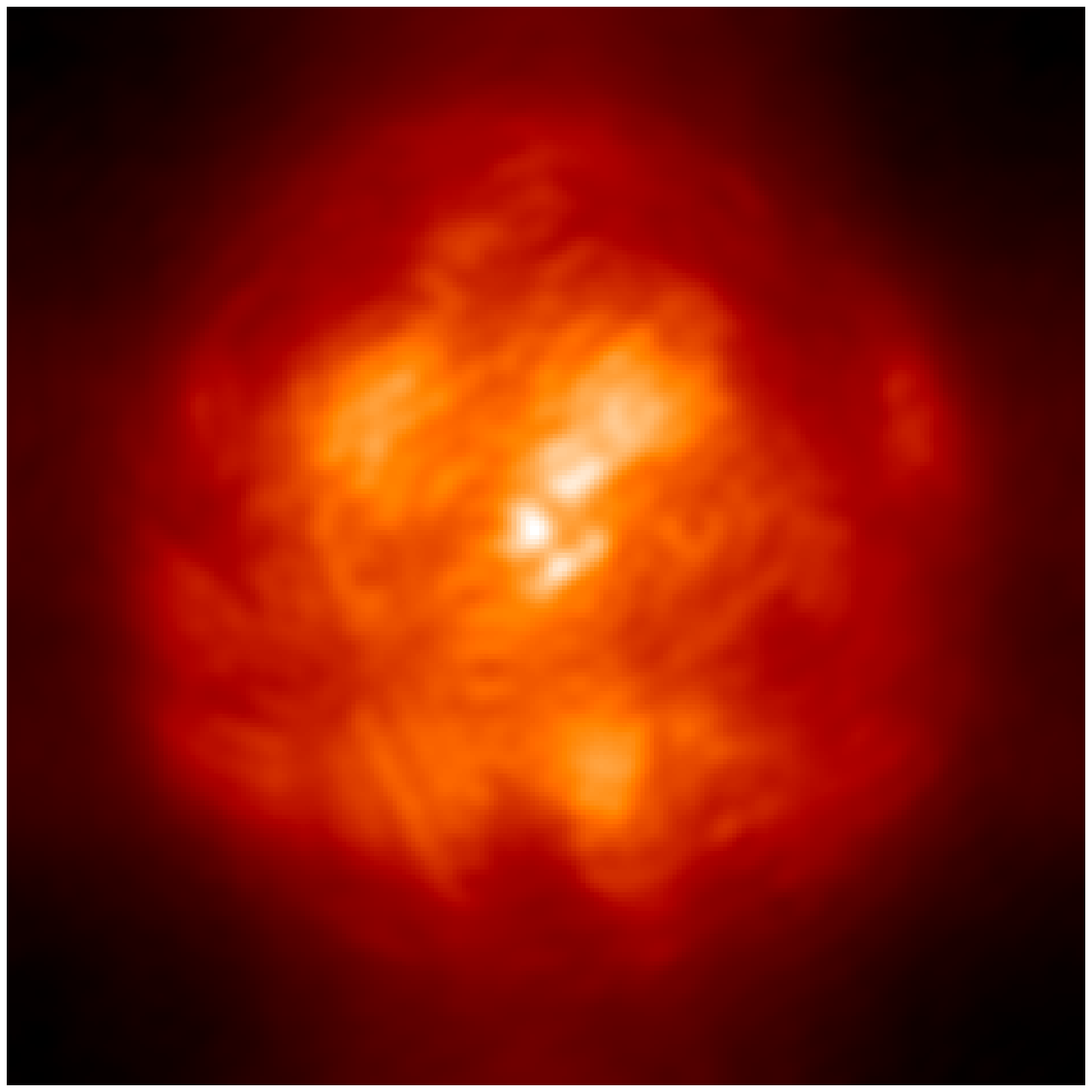} &
\includegraphics[width=39mm]{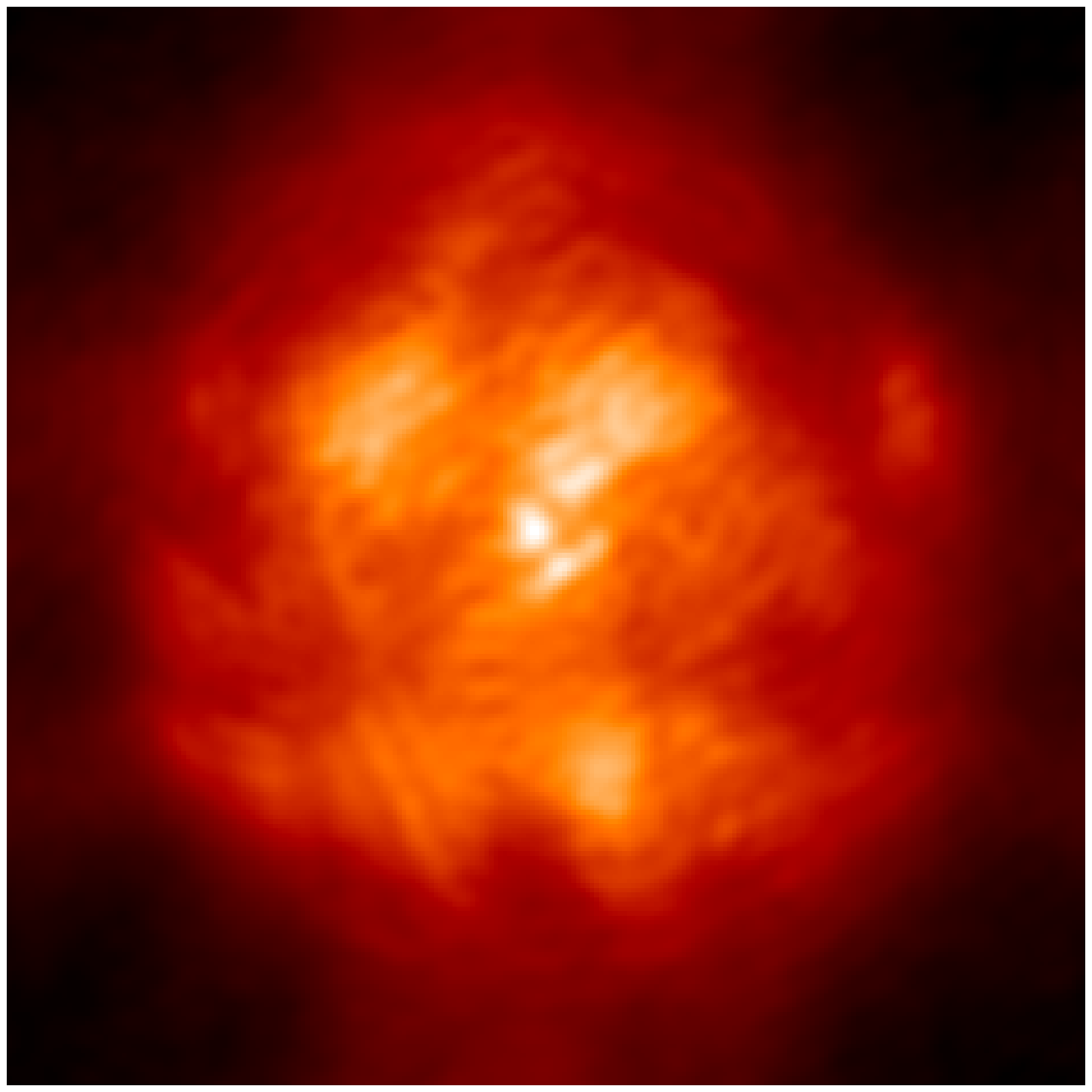} &
\includegraphics[width=39mm]{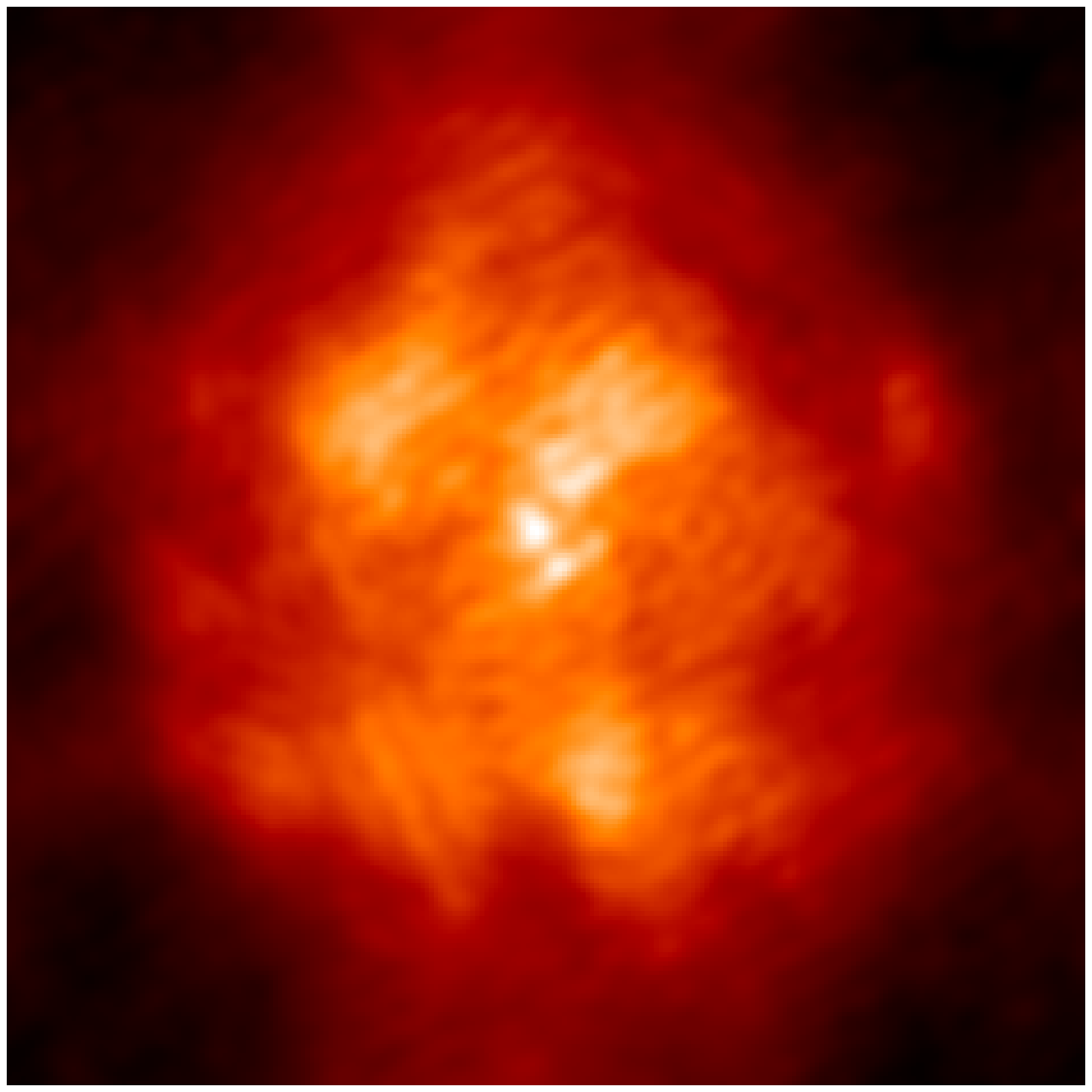} \\
\end{tabular}
\end{center}
\caption[] 
{Direct images of Betelgeuse  with the OVLA configuration of 100 telescopes as a function of the photometric fluctuation between the beams: 10\%, 50\%, 70\%, 90\% of maximum flux loss.\\
\label{fluctu_photom} }
\end{figure}

\begin{figure}[!t] 
\begin{center}
\begin{tabular}{cccc}
\includegraphics[height=60mm]{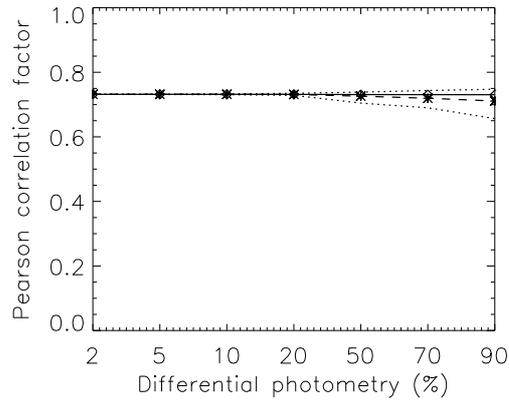}
\end{tabular}
\end{center}
\caption[] 
{Pearson correlation factor as a function of the photometric fluctuations between the beams expressed in pourcents of maximum flux loss. Same legend as Fig. \ref{var_correl_simuOVLA100vsOPD}.\\
\\
\label{var_correl_fluctu_photom} }
\end{figure}

Figures \ref{fluctu_photom} and \ref{var_correl_fluctu_photom} show that the effect of the photometric fluctuations between the beams is not drastic for the hypertelescope. A differential photometry up to 50\% does not really affect the image quality. 
A differential photometry up to 90\% will generate a halo in the image but will preserve the structures in the image.
This specification has also been demonstrated on a fibered testbed \cite{Patru2008}.\\

\subsection{Effect of the photon noise, the read-out noise and the background noise} \label{}

\begin{figure}[!t] 
\begin{center}
\begin{tabular}{cccc}
\includegraphics[width=39mm]{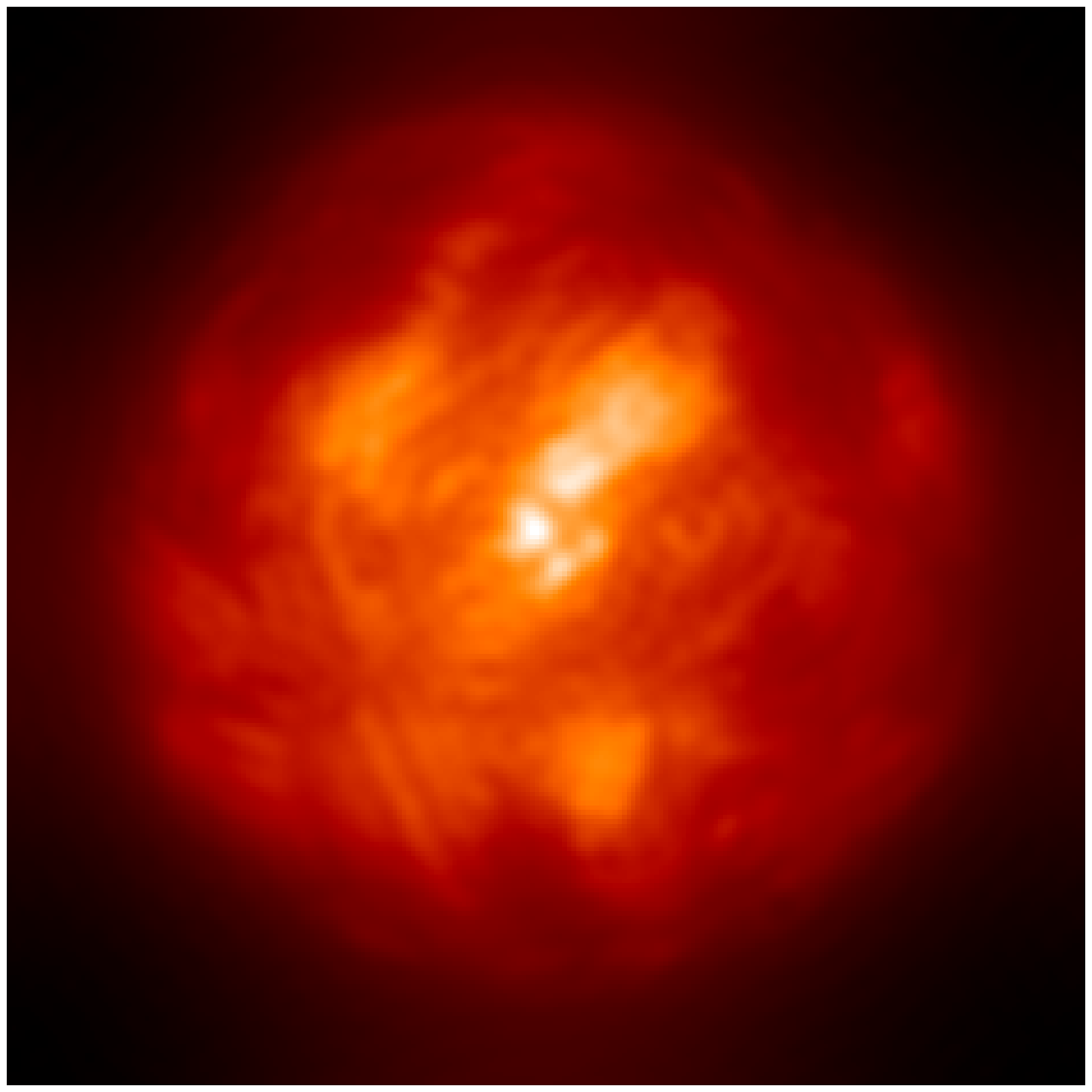} &
\includegraphics[width=39mm]{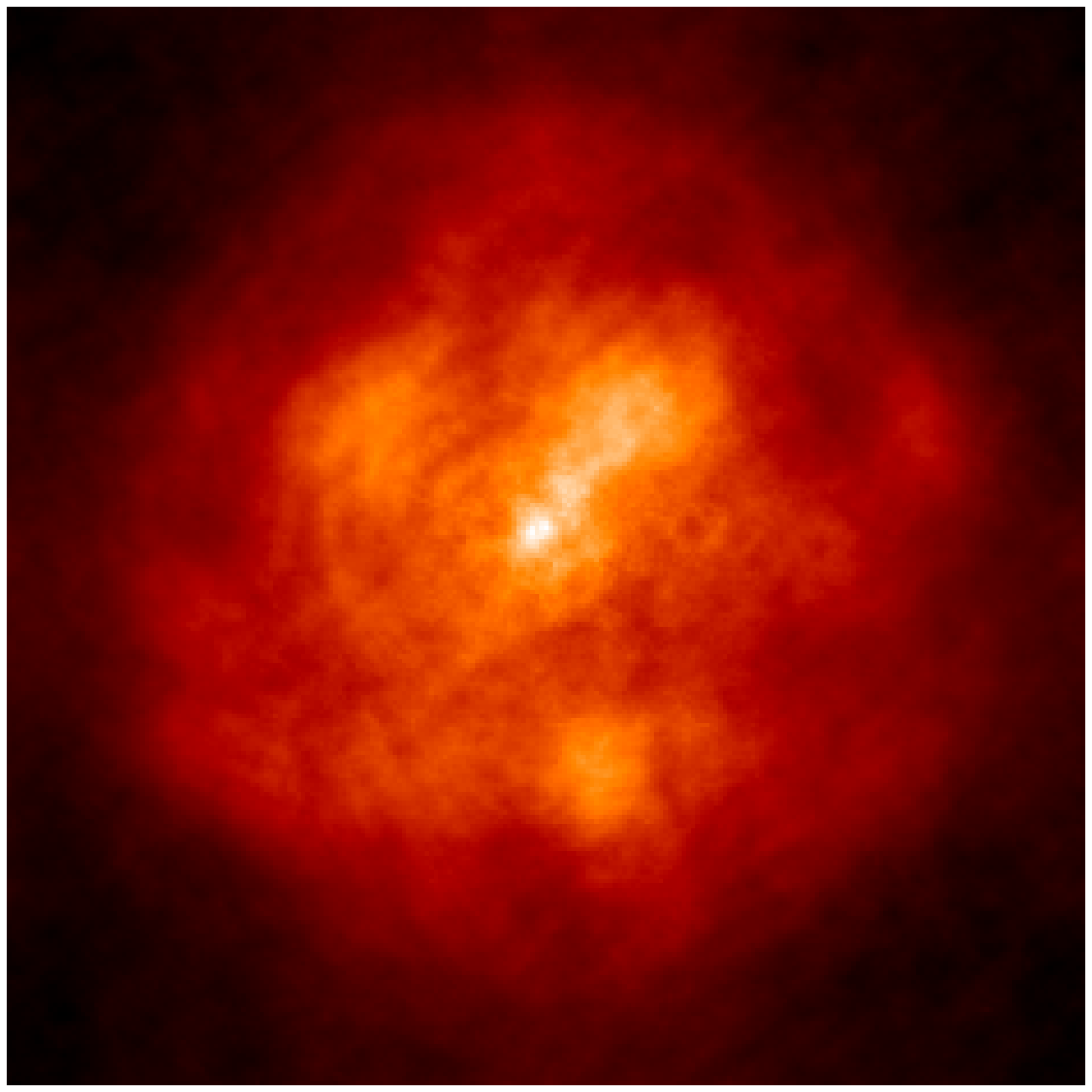} &
\includegraphics[width=39mm]{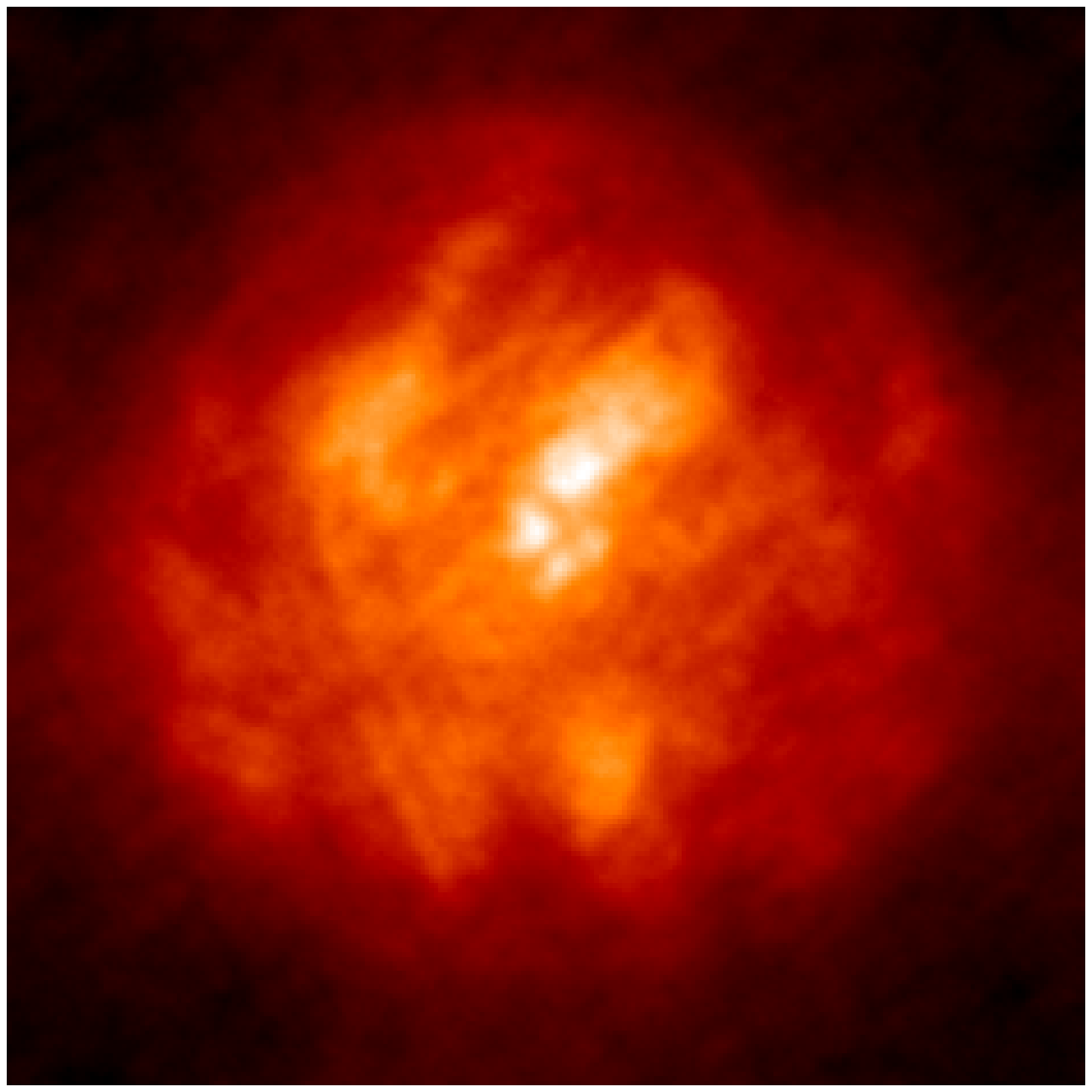} \\
\end{tabular}
\end{center}
\caption[] 
{Direct images of Betelgeuse with the OVLA configuration of 100 telescopes as a function of the read-out noise with a signal to noise ratio of 10 (left), as a function of the photon noise with 2 photons/pixel (middle) and as a function of a background Poisson noise with 10 photons/pixel (right).\\
\label{noise_photon} }
\end{figure}

\begin{figure}[!t] 
\begin{center}
\begin{tabular}{cccc}
\includegraphics[height=42mm]{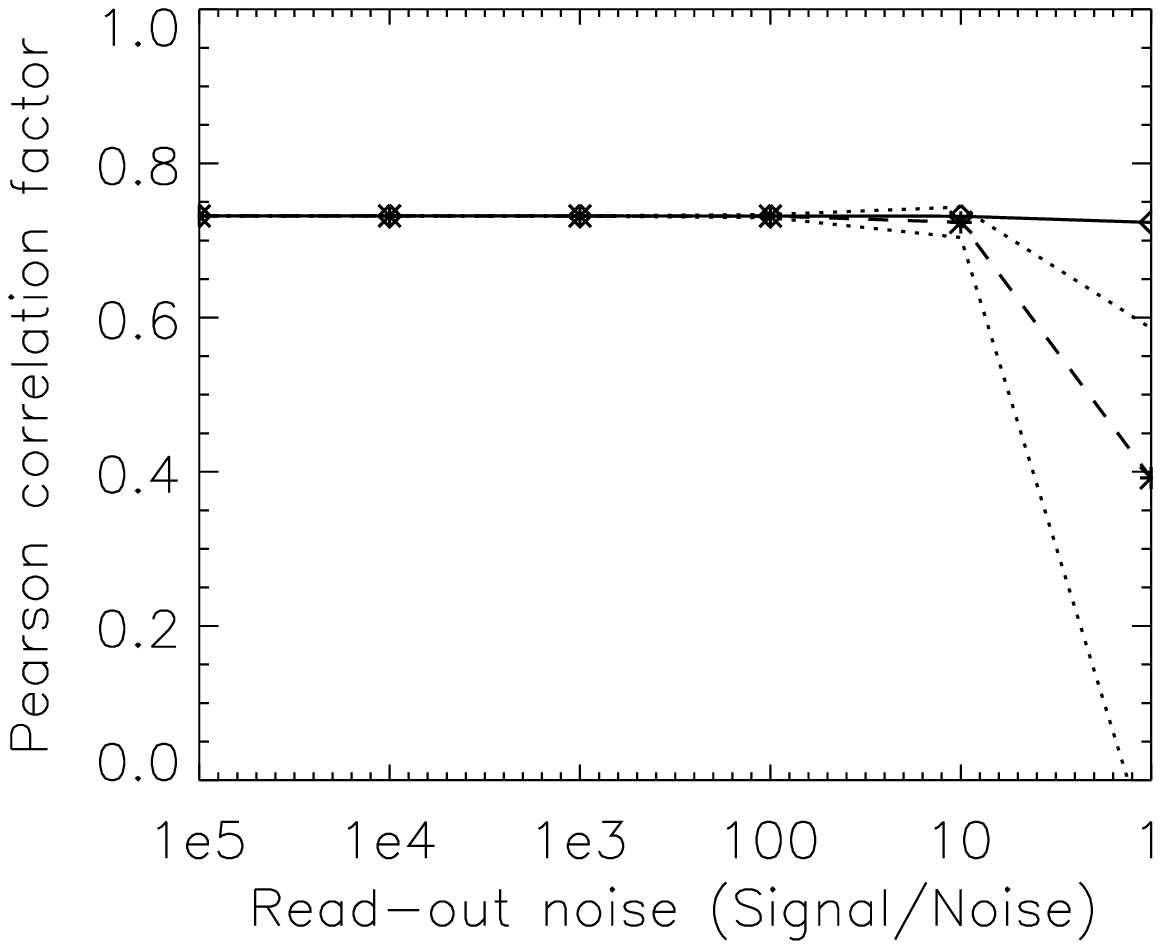} &
\includegraphics[height=42mm]{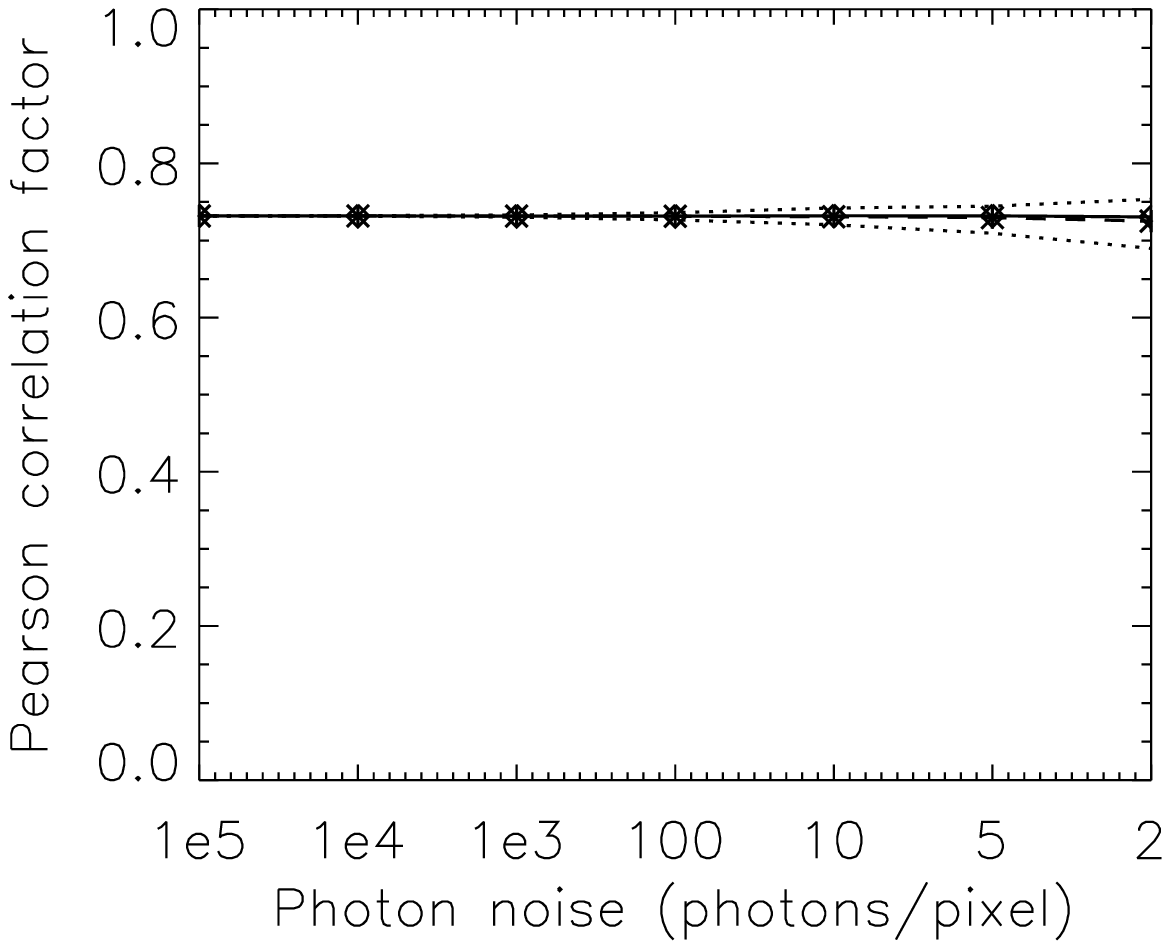} &
\includegraphics[height=42mm]{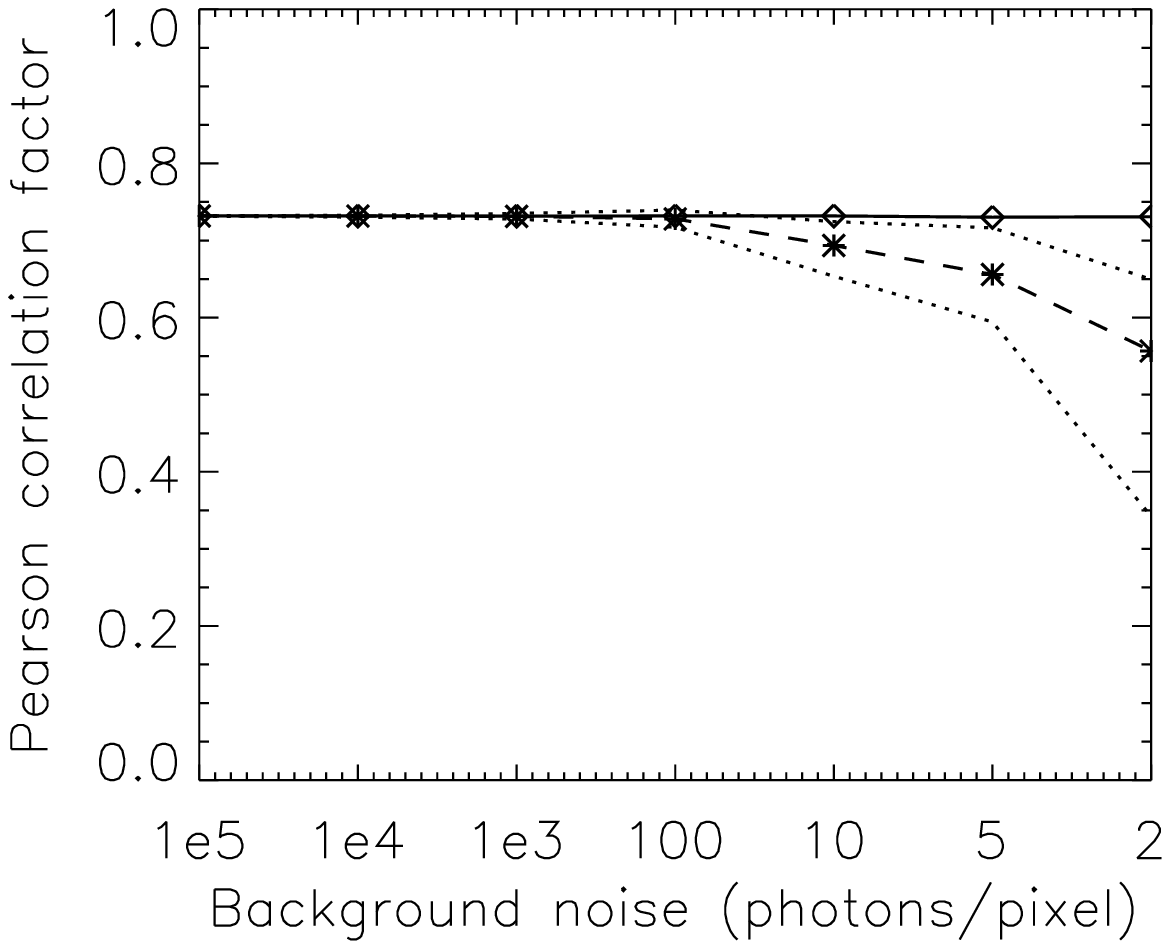} \\
\end{tabular}
\end{center}
\caption[] 
{Pearson correlation as a function of the read-out noise (left), the photon noise (middle) and the background noise (right). Same legend as Fig. \ref{var_correl_simuOVLA100vsOPD}.\\
\\
\label{var_correl_noise} }
\end{figure}

Figures \ref{noise_photon} and \ref{var_correl_noise} show the effect of different kind of noise in the image.
A read-out noise is not critical as long as the signal to noise ratio remains higher than 10. The image becomes sensitive to the photon noise with a signal lower than 10 photons per pixel. The image seems to be more sensitive to a background noise with a poisson signal lower than 100 pixel per pixel, but the thermal background and the zodiacal background remain negligible in the visible wavelength.\\

\section{Plan for image reconstruction} \label{}

A more quantitative analysis of these images will necessarily rely on a more evolved process, involving image reconstruction techniques and taking into account the a priori knowledge of the PSF and/or the multi-chromatic information of the direct images recovered at different wavelength.

A pseudo-deconvolution technique can be thought of by using the approximated formulation of the direct image as the convolution of the object by the interference function multiplied by the diffraction envelope \cite{Labeyrie1996}.
A first idea is that the diffraction envelope is well known (especially in the fibered case where the envelope is fixed and imposed by the fiber) and can be extracted as a flat-field, but the edge of the image is degraded compared to the center part, depending on the signal to noise ratio.
However, the problem is that the strict convolution relationship is lost. The image and the PSF are in fact partially truncated,
which is a problem for the classical methods of deconvolution. To overcome this problem, an hybrid method, based on likelihood maximization, reconstructing simultaneously the object and the PSF has been proposed \cite{Aristidi2006}.

An other way is to use the Fourier information as currently done in interferometry for aperture synthesis imaging. 
Knowing that the hypertelescope concept is a generalization to N sub-apertures of the classical Michelson stellar interferometer with 2 telescopes, it has already been shown that for incoming monochromatic light, the only effect of pupil rearrangement is a shift of the high spatial frequency components of the object spectrum and that this shift can be easily corrected \cite{Tallon1992}.
Note that the spatial frequency peaks will be superimposed in the maximum densification case so that a partial densification ($\gamma<\gamma_{max}/2$) and a non-redundant array are mandatory to preserve the Fourier information.
Moreover, image reconstruction can benefit from algorithms modelling the image to be reconstructed as a sparse combination of vectors in orthogonal representation bases like wavelets. A proposed approach treats the image reconstruction as a sparse approximation problem to include efficient and flexible prior geometric information for the reconstruction \cite{Mary}.\\
\\

\section{Conclusion} \label{}

The astrophysical objectives and the array configuration are strongly dependent to design a hypertelescope.
For a given number of telescopes, there is a trade-off between the resolution, the useful field of view and the contrast in the image \cite{Patru2009}.
In a same way, for a given field of view and a given resolution (chosen as a function of the size of the object), there is a trade-off between the array geometry and the number of required telescopes to optimize either the (u,v) coverage (to recover the intensity distribution) or the dynamic range (to recover the intensity contrast).

To obtain direct images of Betelgeuse with a hypertelescope, a regular layout of telescopes (like KEOPS) is the best array configuration to recover the intensity contrast and the distribution of the granulation cells but it requires a large number of telescopes (several hundreds or thousands). An annular configuration (like OVLA) allows a reasonable number of telescopes (lower than one hundred) to recover the spatial structures but it provides a low-contrast image. The large scale structure remains accessible with a limited number of telescopes (few tens) but the small granulation cells needs in any case many telescopes.

To compensate the huge number of telescopes required in our simulations, a hypertelescope with few apertures can benefit from the earth rotation \cite{Carbillet2002} and from the ability to re-configure telescope locations, although spotted rotated stars generally change in time too quickly to make this last solution efficient. Spectral dispersion can also be helpful to increase the (u, v) coverage by wavelength synthesis \cite{Berger2010}. These aspects should be explore in future studies.

The snapshot images provided by a hypertelescope are highly suitable to follow the evolution of the short timescale convective motions. The snapshot images allow also to feed the entrance plane of an integral field spectrometer to combine low, medium and/or high spectral resolution with the spatial resolution. But a relevant evaluation of the performance of the hypertelescope to study the complex astrophysical phenomena of the red supergiants requires anyway the development of a post-processing of the densified image.

Finally, the photometric fluctuations are not critical ($\Delta$ photometry $<50\%$) contrary to the residual piston requirements (OPD $<\lambda/8$) which requires the development of an efficient cophasing system to fully exploit the imaging capability of a hypertelecope.\\
\\
\\
\\
\\
\\

\bibliography{biblist}   
\bibliographystyle{spiebib}   

\end{document}